\documentclass[a4paper]{article}
\usepackage[utf8]{inputenc}

\usepackage[american]{babel}
\usepackage{amsmath,amstext,amsfonts,amssymb,amsthm,xfrac}
\usepackage{paralist}
\usepackage{url,xspace}
\usepackage[ruled,vlined,fillcomment,noend,linesnumbered]{algorithm2e}
\usepackage{ifthen}
\usepackage{subfig}
\usepackage{graphicx}
\usepackage{fullpage}
\usepackage[shadow]{todonotes}

\usepackage{figlatex}

\usepackage{tikz-timing}
\usetikztiminglibrary[new={char=Q,reset char=R}]{counters}

\newcommand{\period}{\ensuremath{T}\xspace}
\newcommand{\T}{T}

\newcommand{\M}{M}
\newcommand{\ttopt}{\ensuremath{T_{\text{opt}}}\xspace}
\newcommand{\Topt}[1][]{\ensuremath{T_{#1} }\xspace}
\newcommand{\C}{C\xspace}
\newcommand{\D}{D\xspace}
\newcommand{\R}{R\xspace}
\newcommand{\p}{p\xspace}

\newcommand{\recall}{\ensuremath{r}\xspace}
\newcommand{\precision}{\ensuremath{p}\xspace}
\newcommand{\trust}{\ensuremath{q}\xspace}
\newcommand{\muP}{\ensuremath{\mu_{P}}\xspace}
\newcommand{\muNP}{\ensuremath{\mu_{NP}}\xspace}
\newcommand{\munew}{\ensuremath{\mu_e}\xspace}
\newcommand{\waste}{\ensuremath{\textsc{Waste}}\xspace}
\newcommand{\wasteopt}{\ensuremath{\textsc{Waste}_{\text{opt}}}\xspace}
\newcommand{\wasteY}{\ensuremath{\textsc{Waste}_Y}\xspace}

\newcommand{\opt}{\ensuremath{\text{opt}}}
\newcommand{\extr}{\ensuremath{\text{extr}}}

\newcommand{\Tnp}{\ensuremath{T_{\text{R}}}\xspace}
\newcommand{\Tp}{\ensuremath{T_{\text{P}}}\xspace}
\newcommand{\Ty}{\ensuremath{T_{\text{Y}}}\xspace}
\newcommand{\Tlost}[1][]{\ensuremath{T_{\text{lost}#1}}\xspace}
\newcommand{\I}{\ensuremath{I}\xspace}
\newcommand{\MI}{\ensuremath{I'}\xspace}
\newcommand{\EIf}{\ensuremath{\mathbb{E}_{I}^{(f)}}\xspace}

\newcommand{\Textr}[1][q]{\ensuremath{T_{\extr}^{\{#1\}}}\xspace}
\newcommand{\Wregular}{\ensuremath{W_{\mathit{reg}}}\xspace}

\newcommand{\Instant}{\textsc{Instant}\xspace}
\newcommand{\Nockpt}{\textsc{NoCkptI}\xspace}
\newcommand{\Withckpt}{\textsc{WithCkptI}\xspace}
\newcommand{\young}{\textsc{Young}\xspace}
\newcommand{\optexp}{\textsc{OptExp}\xspace}
\newcommand{\ExactPrediction}{\textsc{ExactPrediction}\xspace}
\newcommand{\bestper}{\textsc{BestPeriod}\xspace}

\newcommand{\Young}{\young}
\newcommand{\periodicwithNocheckpointduringI}{\Nockpt}
\newcommand{\periodicwithcheckpointduringI}{\Withckpt}
\newcommand{\periodicignoringI}{\Instant}
\newcommand{\bestyoung}{\bestper \young\xspace}

\newcommand{\bestperiodicwithNocheckpointduringI}{\bestper \Nockpt\xspace}
\newcommand{\bestperiodicwithcheckpointduringI}{\bestper\Withckpt\xspace}
\newcommand{\bestperiodicignoringI}{\bestper\Instant \xspace}
\newcommand{\bestperiodicExactprediction}{\bestper \ExactPrediction\xspace}
\newcommand{\periodicExactprediction}{\ExactPrediction\xspace}

\theoremstyle{definition}

\newcommand{\faultbis}[1]{
\draw[<-, color=red] (#1) -- ($(#1)+(0.2,1.2)$) -- ($(#1)+(0.1,1.4)$) --  ($(#1)+(0.2,2)$) node[above, left] {\scriptsize{failure}};
} 
\newcommand{\predfault}[1]{
\draw[<-, color=green] ($(#1) + (0,1)$) -- ($(#1)+(0.2,1.6)$)  -- ($(#1)+(0.1,1.7)$) -- ($(#1)+(0.2,2)$)  node[above, right] {\scriptsize{Predicted failure}};
} 
\newcommand{\predfaultint}[2]{
\draw[thick, color=green,<->] ($(#1)+(0,1.10)$) -- ($(#1)+(#2,1.10)$) node[above=-0.5pt, midway] {\scriptsize{\I}};
\draw[dashed, color=green] ($(#1)$) --  ($(#1)+(0,1.10)$);
\draw[dashed, color=green] ($(#1)+(#2,0)$) --  ($(#1)+(#2,1.10)$);
} 

\newcommand{\legende}[3]{
\draw[thick, <->] ($(#1)+(0,-0.20)$) -- ($(#1)+(#2,-0.20)$) node[below=-0.5pt, midway] {\scriptsize{#3}};
}

\newcommand{\legendelongue}[4]{
\draw[thick,<->] ($(#1)+(0,-0.20)$) -- ($(#1)+(#2,-0.20)$) node[below=-0.5pt, midway] {\scriptsize{#3}};
\draw[draw=none] ($(#1)+(0,-0.80)$) -- ($(#1)+(#2,-0.80)$) node[below=-0pt, midway] {\scriptsize{#4}};
}

\newcommand{\arrowtime}[2]{
\draw[thick, color=black,->] (0,#1) -- (#2,#1) node[below=-0.5pt, ] {\scriptsize{Time}};
}

\newcommand{\promode}[3]{ 
\draw[thick, color=blue] ($(#1,#3)$) --  ($(#1,#3-1.7)$);
\draw[thick, color=blue] ($(#2,#3)$) --  ($(#2,#3-1.7)$);
\draw[draw=none] ($(#1 ,#3-1.3)$) -- ($(#2,#3-1.3)$) node[midway,color=blue] {\scriptsize{Proactive mode}};
}

\newcommand{\regmode}[3]{
\draw[thick, color=blue] ($(#1,#3)$) --  ($(#1,#3-1.7)$);
\draw[thick, color=blue] ($(#2,#3)$) --  ($(#2,#3-1.7)$);
\draw[draw=none] ($(#1 ,#3-1.3)$) -- ($(#2,#3-1.3)$) node[midway,color=blue] {\scriptsize{Regular mode}};
}
\newcommand{\ttrd}{4} 

\title{Impact of fault prediction on checkpointing strategies}

\author{Guillaume Aupy$^{1}$,Yves Robert$^{1,2}$, Fr\'ed\'eric Vivien$^{1}$ and Dounia Zaidouni$^{1}$\\
 $1.$ Ecole Normale Sup\'erieure de Lyon \& INRIA, France\\
 \url{{Guillaume.Aupy | Yves.Robert | Frederic.Vivien | Dounia.Zaidouni}@ens-lyon.fr}\\
 $2.$ University of Tennessee Knoxville, USA
}

\begin{document}
\maketitle

\begin{abstract}
This paper deals with the impact of fault prediction techniques on checkpointing strategies.
We extend the classical analysis of Young and Daly in the presence of a fault prediction system,
which is characterized by its recall and its precision, and which provides either exact or window-based 
time predictions. We succeed in deriving the optimal value of the checkpointing period (thereby minimizing the 
waste of resource usage due to checkpoint overhead) in all scenarios. These results allow to analytically assess the
key parameters that impact the performance of  fault predictors at very large scale. In addition, the results 
of this analytical evaluation are nicely corroborated by a comprehensive
set of simulations, thereby demonstrating the validity of the model and the accuracy of the results.
\end{abstract}

\section{Introduction}
\label{sec.intro}

In this paper, we assess the impact of fault prediction techniques on checkpointing strategies.
We assume to have jobs executing on a platform subject to faults,
and we let $\mu$ be the mean time between faults (MTBF) of the platform.
In the absence of fault prediction, the standard approach is to take periodic checkpoints, each of
length \C, every period of duration \period. In steady-state utilization of the platform,
the value \ttopt of \period that minimizes the (expectation of the) waste of resource usage due to checkpointing
is easily approximated
as  $\ttopt = \sqrt{2 \mu\C }$, or $\ttopt = \sqrt{2 (\mu +\R)\C }$ (where \R is the duration of the recovery).
The former expression is the well-known Young's formula~\cite{young74},
while the latter is due to Daly~\cite{daly04}.

Now, when some fault prediction mechanism is available, can we compute a better checkpointing period
to decrease the expected waste? and to what extent? Critical parameters that characterize a fault prediction
system are its recall \recall, which is  the fraction of faults that are indeed predicted, and its precision \precision,
which is the fraction of predictions that are correct (i.e., correspond to actual faults). 
The major objective of this paper  is to refine the expression of the
expected waste as a function of these new parameters, and to 
design efficient checkpointing policies that take predictions into account.
We deal with two problem instances, one where the predictor system
provides exact dates for predicted events, and another where it only provides time windows 
during which events take place. 
The key contributions of this paper are the following: (i)
The design of several checkpointing policies, their
analysis, and a new formula for the checkpointing period that extends Young's and Daly's to take predictions into account;
(ii) The analytical characterization of the best policy for each set of parameters; (iii) The validation of the theoretical results via extensive simulations, for both Exponential and Weibull failure distributions; (iv)  The demonstration that even a
poor predictor can lead to a significant reduction of application execution time; and (v)
The demonstration that recall is far more
important than precision, hence giving insight into the design
of future predictors.

The rest of the paper is organized as follows. We first detail the framework in Section~\ref{sec.framework}. We deal with exact date predictions in Section~\ref{sec.no.intervals}, and with time-window based predictions in Section~\ref{sec.intervals}. 
Section~\ref{sec.simulations} is devoted to simulations. Finally, we provide concluding remarks in Section~\ref{sec.conclusion}.

\section{Framework}
\label{sec.framework}

\subsection{Checkpointing strategy}
 
We consider a \emph{platform} subject to faults. Our work is agnostic of the granularity of the platform, which may 
consist either of a single processor, or of several processors that work concurrently and use coordinated checkpointing. 
The key parameter is $\mu$, the mean time between faults (MTBF) of the platform.  If the platform is made of $N$ components
whose individual MTBF is $\mu_{ind}$, then $\mu = \frac{\mu_{ind}}{N}$.
Checkpoints are taken at regular intervals, or periods, of length \period. 
We use \C, \D, and \R for the duration of the checkpoint, downtime and recovery (respectively). 
We must enforce that $\C \leq \period$, and useful work is done only during $\period-\C$ units of time for every period of length \period,
if no fault occurs.
The \emph{waste} due to checkpointing in  a fault-free execution  is $\waste = \frac{\C}{\period}$. 
In the following, the  \emph{waste} always denote the fraction of time that the platform is not doing useful work.

\subsection{Fault predictor}

A fault predictor is a mechanism that is able to predict that some faults will take place, 
either at a certain point in time, or within some time-interval window.
The accuracy of the fault predictor is characterized by two quantities, the \emph{recall}
and the \emph{precision}. The recall \recall is the fraction of faults that are predicted while
the precision \precision is the fraction of fault predictions that are correct.
Traditionally, one defines three types of \emph{events}: (i) \textit{True positive} events are faults that the predictor has been able to predict (let $\textit{True}_P$ be their number); (ii)
\textit{False positive} events are fault predictions that did not materialize as actual faults (let $\textit{False}_P$ be their number);
and (iii)  \textit{False negative} events are faults that were not predicted (let $\textit{False}_N$ be their number).
With these definitions, we have
$\recall = \frac{\textit{True}_P}{\textit{True}_P+\textit{False}_N}$ 
and $\p = \frac{\textit{True}_P}{\textit{True}_P+ \textit{False}_P}$.

\subsection{Fault rates}

In addition to $\mu$, the platform MTBF, 
let $\muP$ be the mean time between predicted events (both true positive and false positive),  and 
let $\muNP$ be the mean time between unpredicted faults (false negative).
Finally, we define the mean time between events as $\munew$ (including all three event types).
The relationships between $\mu$, $\muP$, $\muNP$, and $\munew$ are the following:
\begin{itemize}
\item $\frac{{1-\recall }}{\mu} =  \frac{1}{\muNP}$ (here, $1-\recall$ is the fraction of
  faults that are unpredicted);
   \item $ \frac{\recall}{\mu} =  \frac{\precision}{\muP}$ (here, $\recall$ is the fraction of
  faults that are predicted, and $\precision$ is the fraction of fault predictions that are correct);

\item $\frac{1}{\munew}=\frac{1}{\muP}+\frac{1}{\muNP}$ (here, events are either predicted (true or false), or not).
\end{itemize}

\section{Predictor with exact event dates}
\label{sec.no.intervals}

\begin{figure}
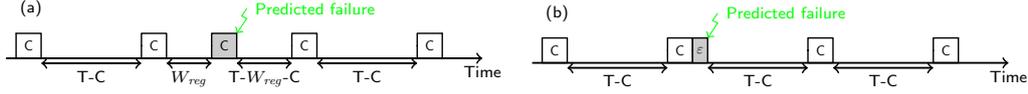

\hspace{-0.5cm}
\scalebox{0.88}{\input{fig/enoughtime.tex}}
\scalebox{0.88}{\input{fig/no_enoughtime.tex}}
\caption{Strategy: (a) Whenever there is enough time to take a checkpoint, the algorithm takes one just before the predicted failure; (b) otherwise, it just executes some extra work.}
\label{fig.enoughtime}
\label{fig.no_enoughtime}
\end{figure}

In this section, we present an analytical model to assess the impact of prediction on periodic checkpointing strategies.
We consider the case
where the predictor is able to provide exact prediction dates, 
and to generate such predictions at least $\C$ seconds in advance, so that a checkpoint can indeed be taken before 
the event (otherwise the prediction cannot be used,
because there is not enough time to take proactive actions).
We consider the following algorithm:\\
(1) While no fault prediction is available, checkpoints are taken
  periodically with period $\period$;\\
(2) When a fault is predicted, we decide whether to take the
  prediction into account or not. This decision is randomly taken: with
  probability \trust, we trust the predictor and take the prediction
  into account, and, with probability $1-\trust$,  we ignore the
  prediction. If we take the prediction into account, there are two cases.
If we have enough time before the prediction date, we take a  checkpoint as late as possible, i.e., so that it completes right
  at the time where the fault is predicted to happen. After the checkpoint, we then complete the execution of the period (see Figure~\ref{fig.enoughtime}(a)). Otherwise, if we do not have enough time
to take an extra checkpoint (because we are already checkpointing), then we do some extra work during $\varepsilon$ seconds (see Figure~\ref{fig.no_enoughtime}(b)). We account for this work as idle time in the expression of the waste, 
to ease the analysis. Our expression of the waste is thus an upper bound.

The rationale for not always trusting the predictor is to avoid taking
useless checkpoints too frequently. Intuitively, the precision $\precision$
of the predictor must be above a given threshold for its usage to be worthwhile.
In other words, if we decide to checkpoint just before a predicted event,
either we will save time by avoiding a costly re-execution if the event does correspond
to an actual fault, or  we will lose time by unduly performing an extra checkpoint.
We need a larger proportion of the former
cases, i.e., a good precision, for the predictor to be really useful.
The following analysis will determine the optimal value of  $\trust$ as a function
of the parameters $\C$, $\mu$, $\recall$, and $\precision$.

\begin{figure*}
\centering
\scalebox{0.9}{\input{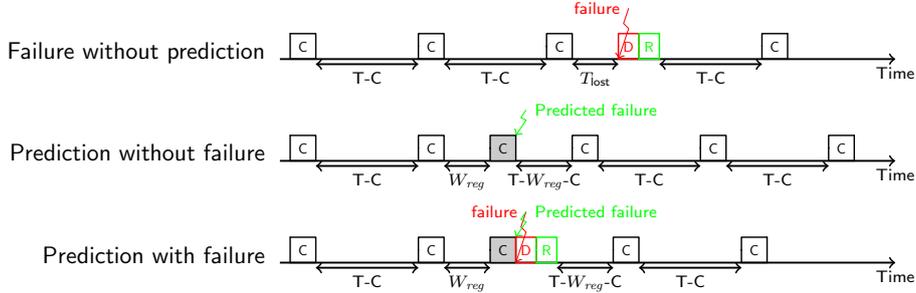}}
\caption{Actions taken when the predictor provides exact dates.}
\label{fig.waste-exact}
\end{figure*}

\subsection{Computing the waste}
\label{sec.nointalg}

Our goal in this section  is to compute a formula for the expected waste. Recall that the waste is the fraction of time that 
the processors do not perform useful computations, 
either because they are checkpointing, or because a failure has struck.
There are four different sources of waste (see Figure~\ref{fig.waste-exact}):\\
(1)  \textbf{Checkpoints:}
During a fault-free execution, the fraction of resources used in checkpointing is $
\frac{\C}{\T}$.\\
(2) \textbf{Unpredicted faults:}
 This overhead occurs each time a unpredicted fault strikes, that is, on average, once every $\muNP$ seconds.
       The time wasted because of the unpredicted fault is then the
      time elapsed between the last checkpoint and the fault, plus
      the downtime and the time needed for the recovery. 
The
      expectation of the time elapsed between the last checkpoint and
      the fault is equal to half the period of checkpoints, because 
the time where the fault hits the system is
      independent of the checkpointing algorithm. 
  Finally, the waste due to unpredicted faults is:  
   $ \frac{1}{\muNP } \left[ \frac{\T}{2} + \D+  \R  \right]$.\\
(3) \textbf{Predictions taken into account:}
Now we have to compute the execution overhead due to a prediction which we trust (hence we checkpoint just before its date). This overhead occurs each time a prediction is made
by the predictor, that is, on average, once every $\muP$ seconds, and that we decide to trust it, with probability $\trust$.
If the predicted event is an actual fault, we waste $\C+\D+\R$  seconds: we waste  $\D + \R $  seconds because the predicted event corresponds to an actual fault, and 
if we have enough time before the prediction date, we waste $\C$ seconds because we take an extra checkpoint as late as possible before the prediction date (see Figure~\ref{fig.enoughtime}(a)). 
Note that if we do not have enough time to take an extra checkpoint (see Figure~\ref{fig.no_enoughtime}(b)), we 
overestimate the waste as $\C$ seconds. If the predicted event is not an actual fault, we waste $\C$
seconds. An actual fault occurs
with probability $\precision$, and a false prediction is made with probability $(1-\precision)$. Averaging with these probabilities, we waste an expected amount of 
 $\left [ \p (\C + \D + \R) + (1-\p) \C \right] $ seconds. Finally, the corresponding overhead is
$\frac{1}{\muP} \trust \left [ \p (\C + \D + \R) + (1-p) \C \right]$.\\
(4) \textbf{Ignored predictions:}
The final source of waste is for predictions that we do not trust. This overhead occurs each time a prediction is made
by the predictor, that is, on average, once every $\muP$ seconds, and that we decide to trust it, with probability $1-\trust$.
 If the predicted event corresponds to an actual fault, we waste  $(\frac{\T}{2} +\D + \R  )$  seconds (as for an unpredicted fault).
Otherwise there is no fault and we took no extra checkpoint, and thus we lose nothing. An actual fault occurs with a probability \precision. 
The corresponding overhead is $\frac{1}{\muP} (1-\trust) \left [  \p (\frac{\T}{2} + \D + \R  ) + (1-\p) 0  \right] $.\\
Summing up the overhead over the four different sources, and after simplification,
we obtain the following equation for the waste:
\begin{equation}
\waste = \frac{\C}{\T} + \frac{1}{\mu} \left[ (1- \recall \trust) \frac{\T}{2} + \D+ \R + \frac{\trust \recall}{\p} \C \right]
\label{eq.waste}
\end{equation}

\subsection{Validity of the analysis}
\label{sec.validity}

Equation~(\ref{eq.waste}) is accurate only when 
two events (an event being a prediction (true or false) or an unpredicted fault) do not take place within the same period.
To ensure that this condition is met with a high probability, we bound the length of the period: 
without predictions, or when predictions are not taken into account,  we enforce the condition $\T < \alpha \mu$; otherwise, with predictions,   we enforce the condition $\T < \alpha \munew$. Here, $\alpha$ is some tuning parameter chosen as follows.
The number of events during a period of length $\T$ can be modeled as a Poisson process of parameter 
$\beta = \frac{\T}{\mu}$ (without prediction) or $\beta = \frac{\T}{\munew}$ (with prediction).
The probability of having $k \geq 0$ faults is $P(X=k) = \frac{\beta^{k}}{k!}  e^{-\beta}$, where $X$ 
is the number of faults. 
Hence the probability of having two or more faults is $\pi = P(X\geq2) = 1 -( P(X=0) + P(X=1)) = 1 - (1+\beta) e^{-\beta}$.
If we assume $\alpha=0.27$ then $\pi \leq 0.03$, 
hence a valid approximation when bounding the period range accordingly. Indeed,
with such a  conservative value for $\alpha$, 
we have overlapping faults for only $3\%$ of the checkpointing segments in average, 
so that the model is quite reliable.

 In addition to the previous constraint, we must always enforce the condition $\C \leq \T$, by construction of the periodic
 checkpointing policy.
Finally, the optimal waste may never exceed $1$; when the waste is equal to $1$, 
the application no longer makes any progress.

\subsection{Waste minimization}
\label{sec.minwaste}

We differentiate twice Equation~\eqref{eq.waste} with respect to \T:
\[ \waste'(\T) = \frac{-\C}{\T^{2}} + \frac{1}{\mu} \left[ (1- \recall \trust) \frac{1}{2}\right] \]
\[ \waste''(\T) = \frac{2 \C}{\T^{3} } > 0 \]
We obtain that $\waste''(\T) $ is strictly positive, hence $\waste(\T) $ is a convex function of $\T$ and admits a unique minimum
on its domain. We also compute $\Textr$, the extremum value of $\T$ that is the unique zero of the function
$\waste'(\T)$,  as $\Textr=\sqrt{ \frac{2 \mu \C}{1-\recall \trust}}$.
Note that this Equation makes sense even when $1-\recall \trust=0$. Indeed this would mean that 
both $\recall=1$ and $\trust=1$: the predictor predicts every fault, and we take proactive action for 
each one of them, there should never be any periodic checkpointing! 
Finally, note that $\Textr$ may well not belong to the admissible domain $[\C, \alpha \munew]$. 

The optimal waste $\wasteopt$ is determined via the following case analysis.
We rewrite the waste as an affine function of $\trust$:
\[ \waste(\trust) = \frac{\recall \trust}{\mu}\left (\frac{\C}{\p}-\frac{\T}{2}\right )+\left ( \frac{\C}{\T}+\frac{\T}{2 \mu}+\frac{\D+ \R}{\mu} \right ) \]
For any value of \T, we deduce that $\waste(\trust)$ is minimized either for $\trust=0$ or for $\trust=1$.
This (somewhat unexpected) conclusion is that the predictor should sometimes be always trusted, 
and sometimes never, but no in-between value
for $\trust$ will do a better job.
Thus we need to minimize the two functions $\waste^{\{0\}}$ and $\waste^{\{1\}}$
over the domain of admissible values for \T, and to retain the best result.

We have $\waste^{\{0\}}(T)= \frac{\C}{\T} + \frac{1}{\mu} \left[ \frac{\T}{2} + \D+ \R  \right]$.
We recognize here the waste function of Young~\cite{young74} and write 
$\wasteY = \frac{\C}{\T} + \frac{1}{\mu} \left[ \frac{\T}{2} + \D+ \R  \right]$.
The function $\wasteY(T)$ is a convex function and reaches its minimum for $\Ty$ in the interval $[\C,\alpha \mu]$:
\begin{itemize}
  \item If ($\C<\Textr[0]<\alpha \mu$):  $\Ty =\Textr[0]=\sqrt{2 \mu \C}$
  \item If ($\Textr[0]<\C$):  $\Ty=\C$
  \item If ($\Textr[0] \geq \alpha \mu$):  $\Ty =\alpha \mu$
   \end{itemize}
Thus, \wasteY ($ = \waste^{\{0\}}$) is minimized for:
\[ \Ty=\min \left ( \alpha \mu,\max(\sqrt{2 \mu \C}, \C )   \right )\]

Similarly, we have:
 $\waste^{\{1\}}(\T)=\frac{\C}{\T} + \frac{1}{\mu} \left[ (1- \recall) \frac{\T}{2} + \D+ \R + \frac{\recall}{\p} \C \right] $.
 The function $\waste^{\{1\}}(T)$ is a convex function and reaches its minimum for \Topt[1] in the interval $[\C,\alpha \munew]$.
\begin{itemize}
  \item If ($\C<\Textr[1]<\alpha \munew$):  $\Topt[1]=\Textr[1]=\sqrt{ \frac{2 \mu \C}{1-\recall}}$
  \item If ($\Textr[1]<\C$):  $\Topt[1]=\C$
  \item If ($\Textr[1] \geq \alpha \munew$):  $\Topt[1]=\alpha \munew$
   \end{itemize}
Thus, $\waste^{\{1\}}$ is minimized for:
\[ \Topt[1]=\min \left ( \alpha \munew,\max(\sqrt{ \frac{2 \mu \C}{1-\recall} }, \C )   \right )\]
Finally, the optimal waste is:
\[ \wasteopt = \min \left (\wasteY(\Ty),\waste^{\{1\}}(\Topt[1]) \right )\]

 \subsection{Prediction and preventive migration}
 \label{sec.migration}

In this section, we make a short digression and 
briefly present an analytical model to assess the impact of prediction and preventive migration on periodic checkpointing strategies. As before, we consider a predictor that is able to predict exactly when faults happen, and to generate these predictions at least  $\C$ 
seconds before the event dates.

The idea of migration consists in moving a task for execution on another node, when a fault
is predicted to happen on the current node in the near future. Note that the faulty node
can later be replaced, in case of a hardware fault, or software rejuvenation can be
used in case of a software fault.
We consider the following algorithm, which is very similar to that used in Section~\ref{sec.nointalg}:
\begin{enumerate}
\item When no fault prediction is available, checkpoints are taken
  periodically with period $\period$.
\item When a fault is predicted, we decide whether to execute the
  migration or not. The decision is a random one: with
  probability \trust we trust the predictor and do the migration and, with probability 1-\trust, we ignore the
  prediction. If we take the prediction into account, we execute the migration as late as possible, so that it completes right
  at the time when the fault is predicted to happen.
\end{enumerate}

As before, we have four different sources of waste. Summing the overhead of the execution of these different sources, we obtain the following equation for the waste (where $\M$ is the duration of a migration):

\begin{subequations}
\begin{align*}
 \waste &=  \frac{\C}{\T} \\
&+  \frac{1}{\muNP } \left[ \frac{\T}{2} + \D + \R  \right] \\
&+  \frac{1}{\muP} \trust \left [ \p (\M) + (1-p) \M \right] \\
&+  \frac{1}{\muP} (1-\trust) \left [  \p (\frac{\T}{2} + \D + \R  ) + (1-\p) 0  \right] 
\end{align*}
\end{subequations}

 After simplification, we get:
\begin{equation}
\waste = \frac{\C}{\T} + \frac{1}{\mu} \left[ (1- \recall \trust) \left (\frac{\T}{2}+\D+ \R \right )  + \frac{\trust \recall}{\p} \M \right]
\label{eq.wasteM}
\end{equation}

Equation~\eqref{eq.wasteM} is very similar to Equation~\eqref{eq.waste}, and the 
minimization of the waste proceeds exactly as in Section~\ref{sec.minwaste}.
 In a nutshell,
$\waste(T) $ is again a convex function and admits a unique minimum over its domain $[\C, \alpha \munew]$,
 the unique zero of the derivative has the same value $\Textr=\sqrt{ \frac{2 \mu \C}{1-\recall \trust}}$,
 and for any value of $T$, the waste is minimized for either $\trust=0$ or $\trust=1$.
 We conduct the very same case analysis as in Section~\ref{sec.minwaste}. 
 
\section{Predictor with a prediction window}
	\label{sec.intervals}

In the previous section, we supposed that the predictor was able to predict exactly when faults will
strike. 
Here, we suppose (maybe more realistically) 
that the predictor gives a \emph{prediction window}, that is an interval of time of length \I during 
which the predicted fault is likely to happen. As before in Section~\ref{sec.no.intervals}: (i) We suppose that we 
have enough time to checkpoint before the beginning of the prediction window; and (ii) 
When a prediction is made, we enforce that the scheduling algorithm has the choice either to take or not to take this
prediction into account, with probability \trust.

We start with a description of the strategies that can be used, depending upon the (relative) length \I
of the prediction window. Let us define two \emph{modes} for the scheduling algorithm:\\
\textbf{Regular}: This is the mode used when no fault prediction is available, 
or when a prediction is available but we decide to ignore it (with probability $1-\trust$). 
In regular mode, we use periodic checkpointing with period \Tnp. 
Intuitively, \Tnp corresponds to the checkpointing period $T$ of Section~\ref{sec.no.intervals}.\\
\textbf{Proactive}: This is the mode used when a fault prediction is available and
we decide to trust it, a decision taken with probability \trust.  Consider 
such a trusted prediction made with the prediction window $[t_0,t_0+\I]$.
Several strategies can be envisioned:\\
(1) \Instant, for \emph {Instantaneous--} The first strategy is to ignore the time-window and to execute the same algorithm
      as if the predictor had given an exact date prediction at time $t_{0}$. Just as described in 
      Section~\ref{sec.no.intervals}, the algorithm interrupts the current period (of scheduled length \Tnp),
      checkpoints during the interval $[t_{0}-C,t_{0}]$, 
      and then returns to regular mode: at time $t_{0}$, it resumes the work needed to complete the interrupted period
      of the regular mode.\\
(2) \Nockpt, for \emph{No checkpoint during prediction window--} 
       The second strategy is intended for a short prediction window: instead of ignoring it,
       we acknowledge it, but make the decision not to checkpoint during it. 
       As in the first strategy, the algorithm interrupts the current period (of scheduled length \Tnp),
      and checkpoints during the interval $[t_{0}-C,t_{0}]$. But here, we return to regular mode
       only at time $t_0+\I$, where we resume the work needed to complete the interrupted period of the regular mode.
       During the whole length of the time-window, we execute work without checkpointing, at the risk
       of losing work if a fault indeed strikes. But for a small value of \I, it may not be
       worthwhile to checkpoint during the prediction window (if at all possible, since there is no choice if $\I < C$).\\
(3) \Withckpt, for \emph{With checkpoints during prediction window--} 
       The third strategy is intended for a longer prediction window and assumes that $\C \leq \I$:
       the algorithm interrupts the current period (of scheduled length \Tnp),
      and checkpoints during the interval $[t_{0}-C,t_{0}]$, but now decides
       to take several checkpoints during the prediction window. 
       The period \Tp of these checkpoints in proactive mode
       will presumably be shorter than \Tnp, to take into account the higher fault probability.
       To simplify the presentation, we use an integer number of periods of length \Tp
       within the prediction window. In the following, we analytically compute the optimal
       number of such periods. But we take at least one period here, hence one checkpoint, which implies $C \leq I$.
       We return to regular mode either right after the fault strikes within the time window 
       $[t_0,t_0+\I]$, or at time $t_0+\I$ if no actual fault happens within this window. 
       Then, we resume the work needed to complete the interrupted period of the regular mode.
    The third strategy is the most complex to describe, and the
complete behavior of the scheduling algorithm is shown in Algorithm~\ref{algo.proactive}.

Note that for all strategies, exactly as in Section~\ref{sec.no.intervals},
we insert some additional work for the particular case where there is not enough time to take a checkpoint
before entering proactive mode (because a checkpoint for the regular mode is currently on-going,
see Figure~\ref{fig.no_enoughtime}(b)). We account for this work as idle time in the expression of the 
waste, to ease the analysis. Our expression of the waste is thus an upper bound.

\begin{algorithm}
  \If{fault happens}{After downtime, execute recovery\;
    Enter \emph{regular} mode\;}
  \If{in \emph{proactive} mode for a time greater than or equal to \I\label{algo.proactive.Ilimit}}{Switch
  to \emph{regular} mode}
  \If{Prediction made with interval $[t, t+I]$ \textbf{and} prediction
    taken into account}{%
    Let $t_\C$ be the date of the last checkpoint under
    \emph{regular} mode to start no later than $t-\C$\;
    \If({(enough time for an extra checkpoint)}) {$t_\C+\C < t-\C$}{Take a checkpoint starting at   
     time $t-\C$\label{algo.proactive.addC}}
    \Else ({(no time for the extra checkpoint)}){
      Work in the time interval $[t_\C+\C, t]$\label{algo.proactive.wait}
    }
    $\Wregular \leftarrow \max \left (0, t-\C - (t_\C + \C ) \right )$ \label{algo.proactive.wreg}\;
    Switch to \emph{proactive} mode at time $t$\;
  }
  \While{in \emph{regular} mode and no predictions are made and no faults
    happen}{
    Work for a time \Tnp-\Wregular-\C and then checkpoint\label{algo.proactive.completion}\;
    $\Wregular \leftarrow 0$\;
  }
  \While{in \emph{proactive} mode and no faults happen}{
    Work for a time \Tp-\C and then checkpoint\;
  }
\caption{\Withckpt.\label{algo.proactive}}
\end{algorithm}

\begin{figure*}
\centering
\scalebox{0.9}{\input{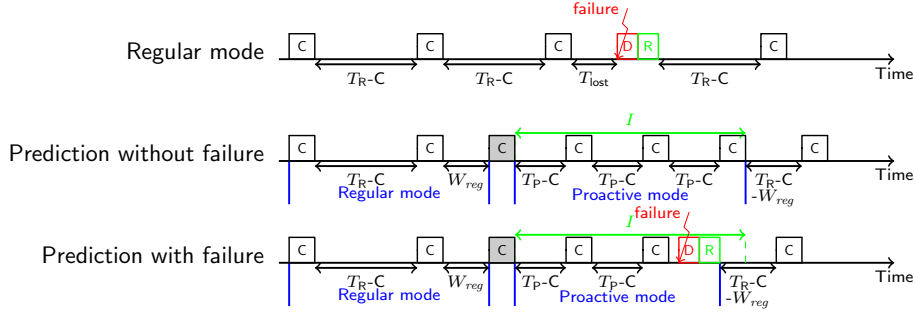}}
	\caption{Outline of Algorithm~\ref{algo.proactive} (strategy \Withckpt). }
\end{figure*}

\subsection{Waste for strategy \Withckpt}
\label{sec-waste-int}

In this section we focus on computing the waste of  \Withckpt, the most complex strategy.
We first compute the fraction of time spent in the \emph{regular} mode (checkpointing with period \Tnp)
and the  fraction of time spent in the \emph{proactive} mode (checkpointing with period \Tp).
Let \MI be the average time spent in the \emph{proactive} mode.
When a prediction is made, we may choose to
ignore it, which happens with probability $1-\trust$. In this case,
the algorithm stays in regular mode and does not spend any time in the proactive mode. With
probability \trust, we may decide to take the prediction into
account. In this case, if the prediction is a false positive event (no actual fault strikes), which happens with probability
$1-\precision$, then the algorithm spends \I units of time in the proactive
mode.  Otherwise, if the prediction is a true positive event (an
actual fault hits the system), which happens with probability
$\precision$, then the algorithm spends an average of $\EIf$ in the
proactive mode. Here $\EIf$ is the expectation of the time elapsed between
the beginning of the prediction window and the time when a fault happens,
knowing that a fault happens in the prediction window. Note that if faults are uniformly distributed across 
the prediction window, then $\EIf = \frac{\I}{2}$. 
Altogether, we obtain $
  \MI   = \trust \left((1-\precision)\I + \precision \EIf\right)$.
Each time there is a prediction, that is, on the average, 
every $\muP$ seconds, the algorithm spends a time $\MI$ in the proactive
mode. Therefore, Algorithm~\ref{algo.proactive} spends a fraction of
time $\frac{\MI}{\muP}$ in the proactive mode, and a fraction of time
$1-\frac{\MI}{\muP}$ in the regular mode. 

As in 
Section~\ref{sec.no.intervals}, we assume that there is a single event of any type (either a prediction (true or false),
or an unpredicted failure) within each interval under study.
The condition $T \leq \alpha \munew$ then becomes
$\Tnp + \I \leq \alpha \munew$, since
$\Tnp+\I$ is the longest time interval considered in the analysis of Algorithm~\ref{algo.proactive}.
We now identify the four different sources of waste, and we analyze
their respective costs:\\
(1) \textbf{Waste due to periodic checkpointing.} There are two cases, depending upon the mode of Algorithm~\ref{algo.proactive}:\\
(a) \textbf{Regular mode.} In this mode, we take periodic checkpoints. 
      We take a checkpoint of
      size \C each time the algorithm has processed work for a time
      $\Tnp-\C$ in the regular mode. This remains true if, after spending
      some time in the regular mode, the algorithm switches to the
      proactive mode,  and later switches back to the regular mode. This
      behavior is enforced by recording the amount of work performed under the regular mode (variable
      \Wregular, at line~\ref{algo.proactive.wreg} of Algorithm~\ref{algo.proactive}), 
      and by taking this value into account at line~\ref{algo.proactive.completion}.
      Given the fraction of time
      that Algorithm~\ref{algo.proactive} spends in the regular mode, this
      source of waste has a total cost of 
        $\left(1 -\frac{\MI}{\muP}\right)\frac{\C}{\Tnp}$.\\
(b)  \textbf{Proactive mode.} In this mode, we take a checkpoint
      of size \C each time the algorithm has processed work for a time
      $\Tp-\C$.
      If no fault happens while the algorithm is in the proactive
      mode, then the algorithm stays exactly a time \I in this mode
      (thanks to the condition at
      line~\ref{algo.proactive.Ilimit}).
      The waste due to the periodic checkpointing is then exactly $\frac{\C}{\Tp}$ 
      (because \Tp divides \I).
      If a fault happens while the algorithm is in proactive mode,
      then, the expectation of the waste due to the periodic checkpointing is
      upper-bounded by the same quantity
      $ \frac{\C}{\Tp}$
     (this is an over-approximation of the waste in that case).
      Overall, taking into account the fraction of time
      Algorithm~\ref{algo.proactive} is in the proactive mode, the cost of this
      source of waste is $\frac{\MI}{\muP}\frac{\C}{\Tp}$.\\
(2) \textbf{Waste incurred when switching to the proactive mode.}
    Each time we take into account a prediction (which happens with
    probability \trust on average every \muP units of time), we start
    by doing one preliminary checkpoint if we have the time to do so
    (line~\ref{algo.proactive.addC}). If we do not have the time to
    take an additional checkpoint, the algorithm do not do any
    processing for a duration of at most \C
    (line~\ref{algo.proactive.wait}). In both cases, the wasted time
    is at most \C and this happens once every
    $\frac{\muP}{\trust}$. Hence, switching from the regular mode to
    the proactive one induces a waste of at most $\frac{\trust}{\muP}C$.\\
(3) \textbf{Waste due to predicted faults.} Predicted faults
    happen with frequency $\frac{\precision}{\muP}$. As we may choose
    to ignore a prediction, there are still two cases
    depending on the mode of the algorithm at the time of the fault:\\
(a) \textbf{Regular mode.} If the algorithm is in regular mode
      when a predicted fault hits, this means that we have chosen to ignore the prediction, a
      decision  taken with probability $(1-\trust)$.
      The time wasted because of the predicted fault is then the
      time elapsed between the last checkpoint and the fault, plus
      the downtime and the time needed for the recovery. 
The
      expectation of the time elapsed between the last checkpoint and
      the fault is equal to half the period of checkpoints, because 
the time where the fault hits the system is
      independent of the checkpointing algorithm. Therefore, 
      the waste due to predicted faults hitting the system in
      regular mode is $\frac{\precision(1-\trust)}{\muP}\left(\frac{\Tnp}{2}+\D+\R\right)$.\\
(b) \textbf{Proactive mode.} If the algorithm is in proactive
      mode when a fault hits, then we have chosen to take the prediction
      into account, a decision that is taken with probability
      $\trust$.
      The time wasted because of the predicted fault is then, in
      addition to the downtime and the time needed for the recovery,
      the time elapsed between the last checkpoint and the fault or,
      if no checkpoint had already been taken in the proactive mode,
      the time elapsed between the start of the proactive mode and the
      fault. 
      Here, we can no longer assume that the time the
      fault hits the system is independent of the checkpointing
      date. This is because the proactive mode starts exactly at the
      beginning of the prediction window. Let \Tlost denote the
      computation time elapsed between the latest of the beginning of
      the proactive mode and the last checkpoint, and the fault
      date. Then the expectation of \Tlost depends on the distribution
      of the fault date in the prediction window. However, we know that 
      whatever the distribution, $\Tlost \leq \Tp$. Therefore we over approximate the 
      waste in that case by
        $\frac{\trust\precision}{\muP}\left(\Tp+\D+\R\right)$.\\
(4)  \textbf{Waste due to unpredicted faults.} There are again two cases,
    depending upon the mode of the algorithm at the time the
    fault hits the system:\\
(a)  \textbf{Regular mode.} In this mode the work done is
      periodically checkpointed with period \Tnp. The time wasted
      because of an unpredicted fault is then the time elapsed
      between the last checkpoint and the fault, plus the downtime
      and the time needed for the recovery. As before, the expectation of this
      value is $\Tlost = \frac{\Tnp}{2}$.
      An unexpected fault hits the system once every
      $\muNP$ seconds on the average. Taking into account the fraction of the time the
      algorithm is in regular mode, the waste due to unpredicted
      faults hitting the system in regular mode is $\left(1-\frac{\MI}{\muP}\right)\frac{1}{\muNP}\left(\frac{\Tnp}{2}+\D+\R\right)$.\\
(b)  \textbf{Proactive mode.} Because of the assumption that a single event
    takes place within a time-interval, we do not consider the very unlikely case
    where a unpredicted fault strikes during a prediction window.
  This amounts to assume that
      $\frac{\MI}{\muP}\frac{1}{\muNP}(\Tp+\D+\R)$ is negligible.

We gather the expressions of the six different types of waste  and simplify  to
obtain the formula of the overall waste:
\begin{align}
   \waste_{\Withckpt}&= 
  \quad \left(\left (1 -\frac{\MI}{\muP} \right )\frac{1}{\Tnp} +
  \frac{\MI}{\muP}\frac{1}{\Tp}
  + \frac{\trust}{\muP}\right)\C + \frac{\precision(1-\trust)}{\muP}\frac{\Tnp}{2}  \nonumber \\
 & + \frac{\precision \trust}{\muP} \Tp
  +\left (1 -\frac{\MI}{\muP} \right ) \frac{1}{\muNP} \frac{\Tnp}{2}  \nonumber \\
 & + \left(\frac{\precision}{\muP}+\left(1-\frac{\MI}{\muP}\right)\frac{1}{\muNP}\right)\left(\D+\R\right)
  \label{eq.proa.waste}
\end{align}

\subsection{Waste of the other strategies}
\label{sec-waste-other}

The waste of the first strategy (\emph{Instantaneous}) is very close to the one given in Equation~\eqref{eq.waste}. 
The difference lies in \Tlost, the expectation of the work lost when a fault is predicted and the 
prediction is taken into account. 
When a prediction is taken into account and the predicted event is an actual fault, the waste in 
Equation~\eqref{eq.waste} was $\frac{\trust \p}{\muP}(\C + \D + \R)$. Because the prediction was exact, \Tlost was equal to 0.
However in our new Equation, the waste for this part is now 
$\frac{\trust \p}{\muP}(\C + \Tlost + \D + \R)$. On average, the fault occurs after a time \EIf.
However, because we do not know the relation between \EIf and \Tnp, then \Tlost has expectation  $\frac{\Tnp}{2}$
 if $\frac{\Tnp}{2} \leq \EIf$. The new waste is then:
\begin{align}
 \waste_{\Instant} = \frac{\C}{\Tnp} + \frac{1}{\mu} \left[ (1- \recall \trust) \frac{\Tnp}{2} + \D+ \R \right.  
 \left. + \frac{\trust \recall}{\p} \C +\trust \recall \min \left ( \EIf, \frac{\Tnp}{2} \right) \right] 
 \label{eq.waste-instant}
\end{align}

As for the second strategy (\emph{No checkpoint during prediction window}), 
we do no longer incur the waste due to checkpointing in proactive mode as we no longer checkpoint in 
proactive mode.
Furthermore, the value of \Tlost in proactive mode becomes \EIf instead of \Tp.
Consequently, the total waste when there is no checkpoint during the proactive mode is:

\begin{align*}
	\waste_{\text{noCkpt}} &=\left (1 -\frac{\MI}{\muP} \right )\frac{\C}{\Tnp}
  + \frac{\trust}{\muP}\C 
  + \frac{\precision(1-\trust)}{\muP}\left (\frac{\Tnp}{2} + \D +\R
  \right) \\
  & + \frac{\precision \trust}{\muP} \left (\EIf + \D
    +\R \right) 
  +\left (1 -\frac{\MI}{\muP} \right ) \frac{1}{\muNP} \left( \frac{\Tnp}{2} + \D +\R \right) \nonumber \\
\end{align*}  

which we rewrite as
\begin{align}
 \waste_{\Nockpt} &=\left(\left (1 -\frac{\MI}{\muP} \right )\frac{1}{\Tnp} +  \frac{\trust}{\muP}\right)\C  
 + \frac{\precision(1-\trust)}{\muP}\frac{\Tnp}{2}  \nonumber \\
  & + \frac{\precision \trust}{\muP} \EIf 
  +\left (1 -\frac{\MI}{\muP} \right ) \frac{1}{\muNP} \frac{\Tnp}{2} \nonumber \\
 & + \left(\frac{\precision}{\muP}+\left(1-\frac{\MI}{\muP}\right)\frac{1}{\muNP}\right)\left(\D+\R\right)
\label{eq.proa.noCkpt.waste}
\end{align}

Note that when $\I=0$, \Instant and \Nockpt are identical. Indeed, we have $\EIf =0$ if $\I=0$, and we
check that Equations~\eqref{eq.waste-instant} and~\eqref{eq.proa.noCkpt.waste} are identical in that case.

\subsection{Waste minimization}
\label{sec-opt-int}

In this section we aim at minimizing the waste of the three strategies,
and then we find conditions to characterize which one is better. Recall that : 
$$\MI = \trust \left ( (1 - \precision) \I + \precision \EIf \right ) $$
\noindent
\textbf{\Withckpt.}
In order to compute the optimal value for \Tp, let us find the portion of the waste that depends on \Tp:
\begin{equation*}
 \waste_{\Tp} =  \frac{\recall \trust}{ \mu}\left ( \frac{ (1 - \precision) \I + \precision \EIf }{\precision} \frac{ \C}{\Tp}  + \Tp  \right )
\end{equation*}
As we can see, the optimal value for \Tp is independent from \trust, but also from $\mu$.
The optimal value for \Tp is thus:
\begin{equation}
	\label{tp.opt.int}
\Tp^{\extr}=\sqrt{ \dfrac{(1 - \precision) \I + \precision \EIf }{p} \C}
\end{equation}
However, for our algorithm to be correct, we want $\frac{\I}{\Tp} \in \mathbb{N}$ (the interval \I 
is partitioned in $k$ intervals of length \Tp, for some integer $k$). We choose $\Tp^{\opt}$ equal 
to either $\frac{\I}{\left \lfloor \frac{\I}{\Tp^{\extr}}\right \rfloor}$ or 
$\frac{\I}{\left \lfloor \frac{\I}{\Tp^{\extr}}\right \rfloor +1}$, depending on the value that 
minimizes $\waste_{\Tp}$. Note that we also have the constraint $\Tp^{\opt} \geq \C$, hence 
if both values are lower than \C, then $\Tp^{\opt}=\C$.

Now that we know that $\Tp^{\opt}$ is independent from both \trust and \Tnp, we can see the waste in 
Equation~\eqref{eq.proa.waste} as a function of two variables. 
One can see from Equation~\eqref{eq.proa.waste} that the waste is an affine function of \trust. 
This means that the minimum is always reached for either
$\trust=0$ or $\trust=1$.
We now consider the two functions $\waste_{\text{withCkpt}\{\trust=0\}}$ and $\waste_{\text{withCkpt}\{\trust=1\}}$ in order to 
minimize them with respect to \Tnp. First we have:

\begin{equation}
	\label{waste.int.q0}
	\waste_{\text{withCkpt}\{\trust=0\}} =\frac{\C}{\Tnp} 
		+ \frac{1}{\mu}\left ( \frac{\Tnp}{2} + \D +\R \right ) 
\end{equation}
As expected, this is exactly the equation without prediction, the study of the optimal solution has been 
done in Section~\ref{sec.no.intervals}, it is minimized when $\Tnp^{\opt_0} =\min \left( \alpha \munew - \I, \max \left ( \sqrt{2  \C \mu}, \C \right )\right )$. 

Next we have:
\begin{align}
	\label{waste.int.q1}
	\waste_{\text{withCkpt}\{\trust=1\}} &= \left (1 -\frac{\recall \left ( (1 - \precision) \I + \precision \EIf \right )}{\precision \mu} \right ) \left ( \frac{\C}{\Tnp} +  \frac{1-\recall}{\mu}\frac{\Tnp}{2} \right )\nonumber \\
		& +\frac{\recall}{ \mu}\left ( \frac{\left ( (1 - \precision) \I + \precision \EIf \right )}{\precision} \frac{\C}{\Tp^{\opt}}  + \Tp^{\opt} \right )  + \frac{\recall }{\precision \mu}\C  \nonumber \\ 
    	      & + \left(\frac{\recall}{\mu}+\left (1 -\frac{\recall  \left ( (1 - \precision) \I + \precision \EIf \right )}{\precision \mu} \right )\frac{1-\recall}{\mu}\right)\left(\D+\R\right)
\end{align}
This equation is minimized when 
\[
\Tnp^{\opt_1} = \sqrt{ \dfrac{2 \mu\C }{(1-\recall )}}
\]
One can remark that this value is equal to the result without intervals (Section~\ref{sec.no.intervals}). 
Actually, the only impact of the prediction interval \I is the moment when we should take 
a pre-emptive action.
Note that when $\recall=0$ (this means that there is no prediction),  we have
$\Tnp^{\opt_1} = \Tnp^{\opt_0} $, and we retrieve Young's formula~\cite{young74}.

Finally, we know that the waste is defined for $\C \leq \Tnp \leq \alpha \munew - \I$. Hence, if 
$\Tnp^{\opt_1} \notin [\C,\alpha \munew - \I]$, this solution is not satisfiable. However  
Equation~\eqref{waste.int.q1} is convex, so the optimal solution is \C if $\Tnp^{\opt_1} < \C$, and 
$\alpha \munew - \I$ if $\Tnp^{\opt_1} > \alpha \munew$.
Hence, when $\trust=1$, the optimal solution should be 
\begin{equation}
	\label{tnp.opt.int}
\min \left (\alpha \munew - \I,\max \left (\sqrt{ \dfrac{2 \mu\C }{(1-\recall )}},\C \right )\right).
\end{equation}

\noindent
\textbf{\Instant}. The derivation is similar .
The optimal value for \trust is either $0$ or $1$, thus we consider $\waste_{\Instant}^{\{0\}}  = \wasteY$ and $\waste_{\Instant}^{\{1\}}$.
 If $\EIf>\frac{\Tnp}{2}$, then $\waste_{\Instant}^{\{0\}} < \waste_{\Instant}^{\{1\}}$, so we can assume $\min(\EIf, \frac{\Tnp}{2}) = \EIf$.
Then we derive that
 $\waste_{\Instant}^{\{1\}}$  is minimized for  $\Tnp^{\opt_1}$ as before.\\
\noindent
\textbf{\Nockpt}.
One can see that Equation~\eqref{eq.proa.noCkpt.waste} and Equation~\eqref{eq.proa.waste} only differ by the quantity :
$$\frac{\trust \recall}{\mu}\left ( \frac{(1 - \precision) \I + \precision \EIf}{\precision} \frac{\C}{\Tp^{\opt}}  + \Tp^{\opt} - \EIf \right )$$
This value is linear in \trust and a constant with regards to \Tnp. Hence the minimization is almost 
the same.

Once again we can see that the optimal value for \trust is either 0 or 1. We can consider the 
two functions $\waste_{\text{noCkpt}\{\trust=0\}}$ and $\waste_{\text{noCkpt}\{\trust=1\}}$.
We remark that $\waste_{\text{noCkpt}\{\trust=0\}} = \waste_{\text{withCkpt}\{\trust=0\}}$, and hence that the study has 
already been done.
As for $\waste_{\text{noCkpt}\{\trust=1\}}$, it is also minimized when  
$\Tnp^{\opt} = \sqrt{ \dfrac{2 \mu\C }{(1-\recall )}}$.

Finally, the last step of this study is identical to the previous minimization, and the optimal solution 
when $\trust=1$ is defined by :
$$\Tnp^{\opt_1}=\min \left (\alpha \munew - \I,\max \left (\sqrt{ \dfrac{2 \mu\C }{(1-\recall )}},\C \right )\right)$$

\noindent
\textbf{Summary}. Finally in this section, we consider the waste for the two algorithms that take the prediction window 
into account (the one that does not checkpoint during the prediction window, and the one that 
checkpoints during the prediction window), and try to find conditions of dominance of one strategy 
over the other.
Since the equation of the waste is identical when $\trust=0$, let us consider the case when $\trust=1$.
We have seen that:

	\begin{align}
	\label{diff.waste.algo}
 (\waste_{\text{withCkpt}\{\trust=1\}} - \waste_{\text{noCkpt}\{\trust=1\}}) & = 
 \frac{\recall  \left ( (1 - \precision) \I + \precision \EIf \right )}{\precision \mu}\frac{\C}{\Tp^{\opt}} \nonumber \\
 & + \frac{\recall}{\mu} \left ( \Tp^{\opt} - \EIf \right )
 \end{align}

We want to know when Equation~\eqref{diff.waste.algo} is nonnegative (meaning that it is beneficial 
not to take any checkpoints during  proactive mode). We know that this value is minimized when 
$\Tp^{\extr}=\sqrt{ \dfrac{ (1 - \precision) \I + \precision \EIf}{p}\C}$ (Equation~\eqref{tp.opt.int}),
then a sufficient condition would be to study the equation : $$\waste_{\text{withCkpt}\{\trust=1\}} - \waste_{\text{noCkpt}\{\trust=1\}} \geq 0$$
with $\Tp^{\extr}$ instead of $\Tp^{\opt}$. That is:
\begin{align}
&\frac{\recall  (1 - \precision) \I + \precision \EIf}{\precision \mu}\frac{\C}{\sqrt{ \dfrac{ (1 - \precision) \I + \precision \EIf}{p} \C}} + \frac{\recall}{\mu} \left ( \sqrt{ \dfrac{ (1 - \precision) \I + \precision \EIf }{p} \C} - \EIf\right ) &\geq 0 \nonumber\\
& \Leftrightarrow 2\sqrt{ \dfrac{ (1 - \precision) \I + \precision \EIf}{p} \C} \geq \EIf\!^2
\label{cond.noCkpt}
\end{align}

Consequently, we can say that if Equation~\eqref{cond.noCkpt} is matched, then 
$\waste_{\text{noCkpt}}$ $\leq \waste$, the algorithm where we do not checkpoint during the 
proactive mode has a better solution than Algorithm~\ref{algo.proactive}. For example, if we 
assume that faults strike uniformly during the prediction window $[t_{0}, t_{0}+\I]$, in other words, 
if $0 \leq x \leq \I$, the probability that the fault occurs in the interval $[t_{0}, t_{0}+x]$ is 
$\frac{x}{\I}$, then $\EIf =\frac{I}{2}$, and our condition becomes 
\[
\I \leq 16 \frac{1 - \sfrac{\precision}{2}}{\precision}\C.
\]

We can now finish our study by saying that in order to find the optimal solution, one should compute 
both optimal solutions for $\trust=0$ and $\trust = 1$, for both algorithms, and choose the one that 
minimizes the waste, as was done in Section~\ref{sec.no.intervals}, except when 
Equation~\eqref{cond.noCkpt} is valid, then we can focus on the computation of the waste of 
the algorithms that does not checkpoint during  proactive mode.

\section{Simulation results}
\label{sec.simulations}

In order to validate our model, we have instantiated it with several scenarios. 
The experiments use parameters that are representative of current and forthcoming large-scale platforms~\cite{j116,Ferreira2011}.
We have $C=R=10mn$, and $D=1mn$.
The individual (processor) MTBF $\mu_{ind} = 125$ years, and the total number of processors $N$
varies from $N=16,384$ to $N=524,288$, so that the platform MTBF $\mu$ varies from $\mu=4,000mn$ (about $1.5$ day) down to 
$\mu=125mn$ (about $2$ hours).
For instance the Jaguar platform, with $N=45,208$ processors, is reported to experience about one failure per 
day~\cite{6264677}, which leads 
to $\mu_{ind} = \frac{45,208}{365}\approx 125$ years.

We have analytically computed the optimal value of the waste for each strategy (using the formulas of Section~\ref{sec-opt-int}) 
using a computer algebra software.  In order to check the accuracy of our model, 
we have compared the results with those from simulations using a fault generator.
Our simulation engine generates a random trace of failures, parameterized either by an Exponential failure distribution or by a Weibull distribution law with shape parameter $0.5$ and $0.7$; Exponential failures are widely used for theoretical studies, while Weibull 
failures are representative of
the behavior of real-world platforms~\cite{Weibull1,Weibull2,Heien:2011:MTH:2063384.2063444}. 
With probability \recall, we decide if a failure is predicted or not. In both cases, the distribution is scaled so that its expectation corresponds to the platform MTBF $\mu$.
Then the simulation engine generates another random trace of false predictions (whose distribution is 
identical to the first trace or a uniform distributions). This second distribution is scaled 
so that its expectation is $\frac{\precision \mu}{\recall (1-\precision)}$, the inter-arrival time of false predictions.
Finally, both traces are merged to derive the final trace with all events. Each value reported for the simulations is the average of $100$ randomly generated experiments.

In the simulations, we compare up to ten checkpointing strategies. Here is the list:\\
$\bullet$ \young is the periodic checkpointing strategy of period 
$\Textr[0] = \sqrt{2 \mu \C}$ given in~\cite{young74}. Note that Daly's formula~\cite{daly04} leads to the same results.\\
$\bullet$ \ExactPrediction is derived from the  strategy
Section~\ref{sec.no.intervals} (with exact prediction dates). However, in the simulations, we always take prediction into account
and use an uncapped period $\Textr[1] = \sqrt{ \dfrac{2 \mu\C }{1-\recall }}$ instead of 
$\Topt[1] = \min(\alpha \munew - \I, \max(\C, \Textr[1]))$.\\
$\bullet$ Similarly, \Instant,  \Nockpt and 
 \Withckpt are the three strategies described in Section~\ref{sec.intervals}, with the same modification: 
we always take prediction into account
and use an uncapped period $\Textr[1] $ instead of 
$\Topt[1]$ in regular mode.\\
$\bullet$ To assess the quality of each strategy, we compare it with its \bestper counterpart, defined as 
the same strategy but using the best possible
period $\Tnp$. This latter period is computed via a brute-force numerical 
search for the optimal period.

The rationale for modifying the strategies described in the previous sections is of course to better assess the impact of prediction.
For the computer algebra plots, in addition to the waste with the \emph{capped periods} given
 in Section~\ref{sec-opt-int}, i.e., 
 with $\Topt[0]= \Ty = \min(\alpha \mu, \max(\C, \Textr[0]))$, and $\Topt[1] = \min(\alpha \munew - \I, \max(\C, \Textr[1]))$,
we also report the waste obtained for the \emph{uncapped periods},
i.e., using  $\Topt[0] = \Textr[0]$ without prediction and $\Topt[1] = \Textr[1] = \sqrt{ \dfrac{2 \mu\C }{1-\recall }}$ with 
prediction. The objective is twofold: (i) Assess whether the validity of the model can be extended; and (ii) Provide an exact match
with the simulations, which mimic a real-life execution and do allow for an arbitrary number of faults per period.

\subsection{Predictors from the literature}

\begin{figure*}
\centering
\resizebox{\textwidth}{!}{
\includegraphics[scale=0.5]{fig/simulations/legende.fig}}
\\
\resizebox{\textwidth}{!}{
\subfloat[Capped periods]
{
\includegraphics[scale=0.16]{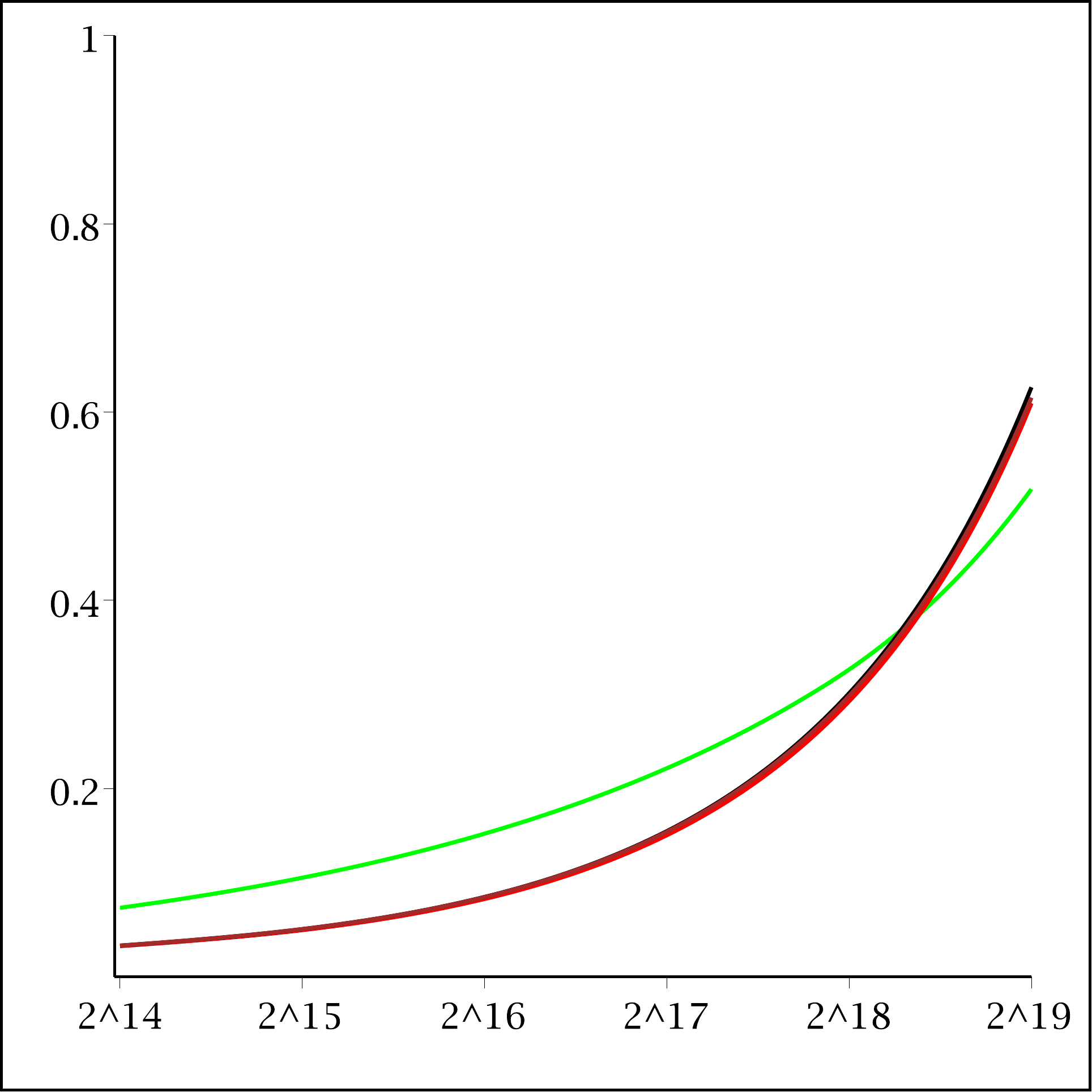}
}
\subfloat[Uncapped periods]
{
\includegraphics[scale=0.16]{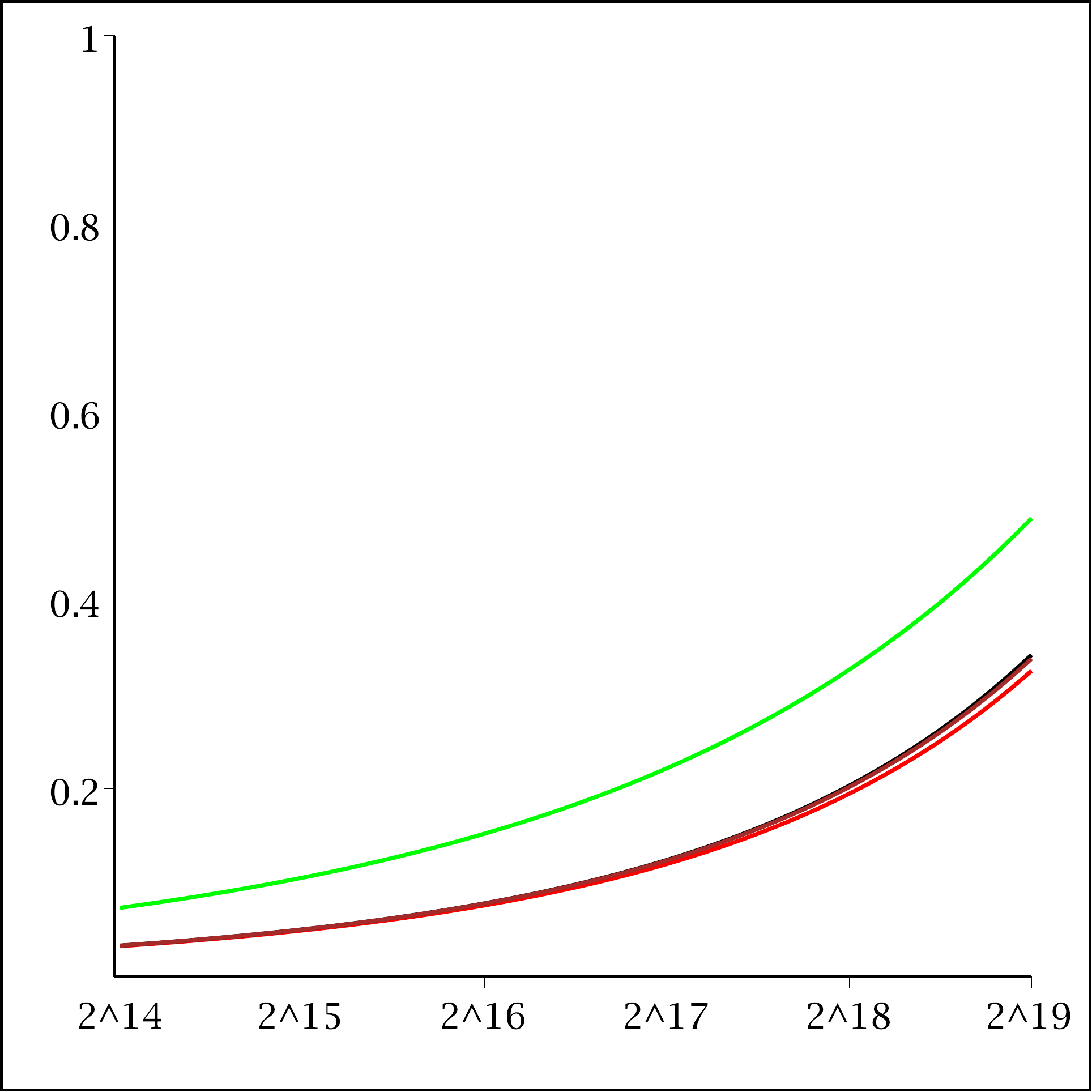}
}
\subfloat[Exponential]
{
\includegraphics[scale=0.36]{fig/simulations/r085p083I300-EXP-fixedC-appli0-platform-variation.fig}
}	
\subfloat[Weilbull $k=0.7$]
{
\includegraphics[scale=0.36]{fig/simulations/r085p083I300-WEIBULL-07-fixedC-appli0-platform-variation.fig}
}
\subfloat[Weilbull $k=0.5$]
{
\includegraphics[scale=0.36]{fig/simulations/r085p083I300-WEIBULL-05-fixedC-appli0-platform-variation.fig}
}
}
\\
\resizebox{\textwidth}{!}{
\subfloat[Capped periods]
{
\includegraphics[scale=0.16]{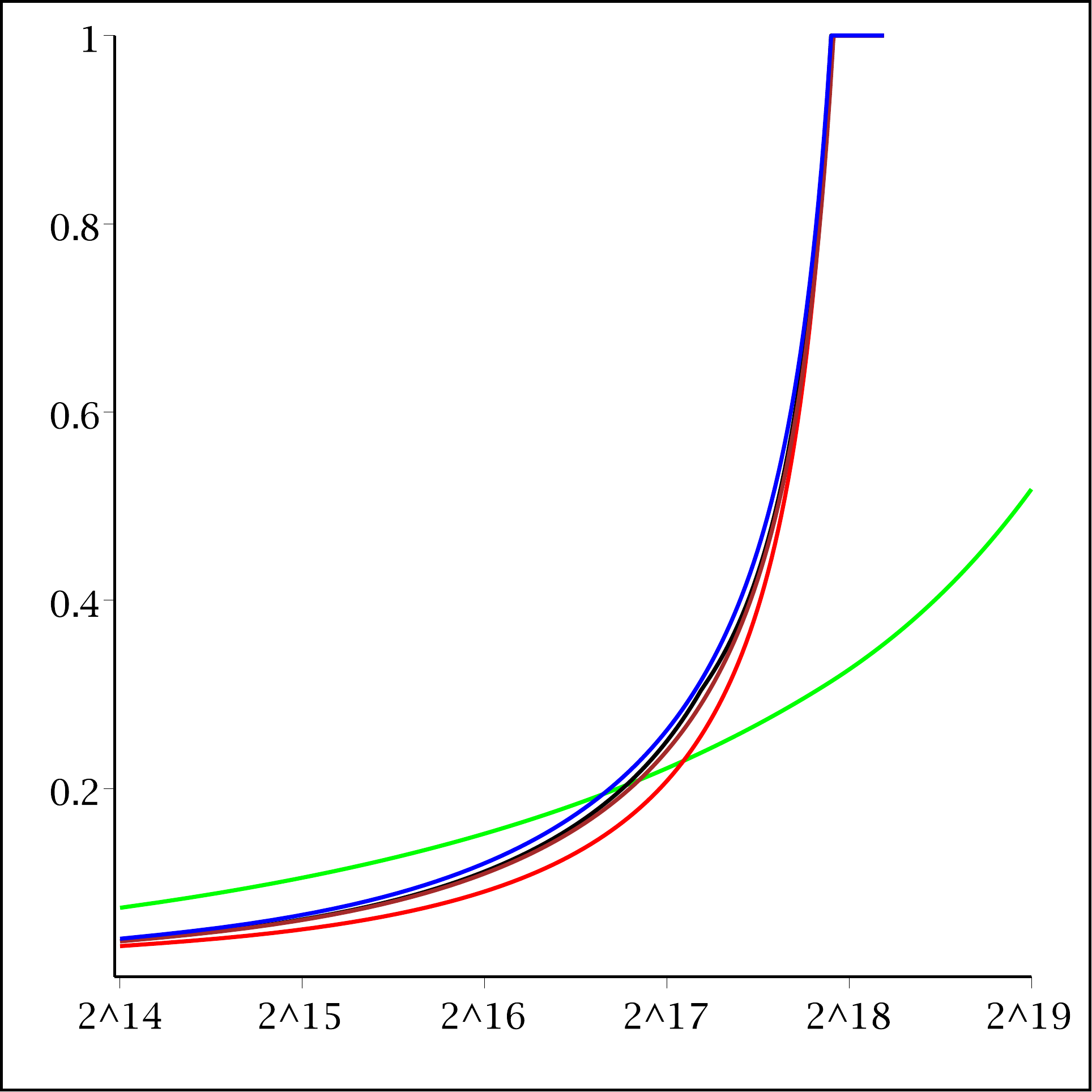}
}
\subfloat[Uncapped periods]
{
\includegraphics[scale=0.16]{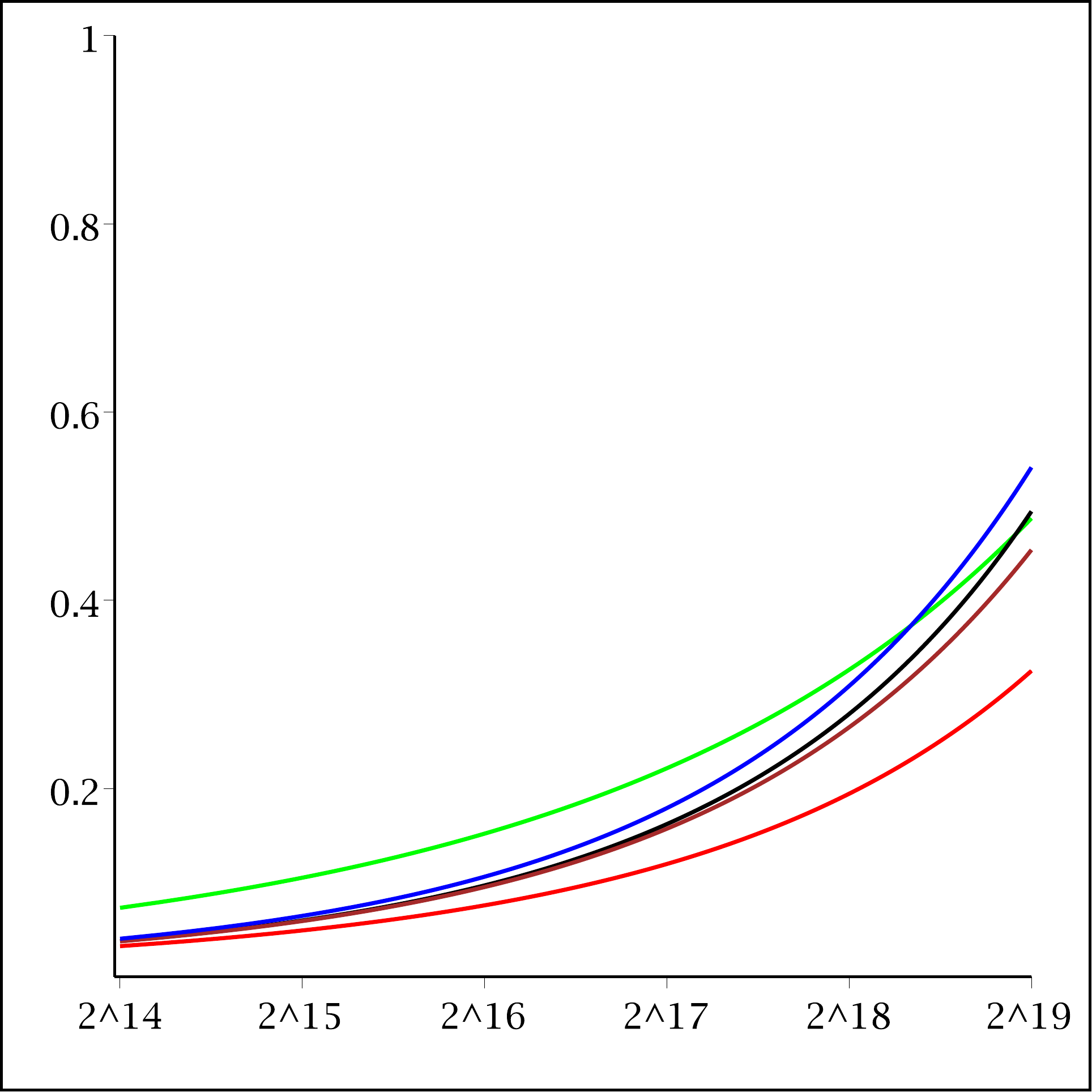}
}
\subfloat[Exponential]
{
\includegraphics[scale=0.36]{fig/simulations/r085p083I3000-EXP-fixedC-appli0-platform-variation.fig}
}	
\subfloat[Weilbull $k=0.7$]
{
\includegraphics[scale=0.36]{fig/simulations/r085p083I3000-WEIBULL-07-fixedC-appli0-platform-variation.fig}
}
\subfloat[Weibull $k=0.5$]
{
\includegraphics[scale=0.36]{fig/simulations/r085p083I3000-WEIBULL-05-fixedC-appli0-platform-variation.fig}
}}
\caption{Waste for the different heuristics, with\,$\precision=0.82$,\,$\recall=0.85$,\,$\I=300$s\,(first row)\,or\,$\I=3,000$s\,(second\,row) 
and with a trace of false predictions parametrized by a distribution identical to the distribution of the trace of failures.}
	\label{fig.082.085}
\end{figure*}

We first experiment with two predictors from the literature: one accurate predictor 
with high recall and precision~\cite{5958823}, namely
with $\precision=0.82$ and $\recall=0.85$, and another predictor with more limited recall 
and precision~\cite{5542627}, namely
with $\precision=0.4$ and $\recall=0.7$. In both cases, we use two different time-windows,
$\I=300s$ and $\I=3,000s$. The former value does not allow for checkpointing within the prediction window, while the
latter values allow for several checkpoints. Note that we always compare the results with \ExactPrediction,  the strategy that
assumes exact prediction dates.
Figures~\ref{fig.082.085} and~\ref{fig.04.07} show the average waste degradation of the
ten heuristics for both predictors, as a function of the number of processors $N$. We draw the plots as a function of the number of processors $N$ rather than of the platform MTBF $\mu = \mu_{ind}/N$ , because it is more natural to see the waste increase with larger platforms; however,  this work is agnostic of the granularity of the processors and intrinsically focuses on the impact of  
the MTBF on the waste. 

The first observation is that the prediction is always useful for the whole set of parameters under study! 
The second observation is the good correspondence between  analytical results and simulations
in Figures~\ref{fig.082.085} and~\ref{fig.04.07}  (compare subfigures (a) and (b) with (c), (d) and (e), and 
subfigures (f) and (g) with  (h), (i) and (j)). This shows the validity of the model for the whole range of distributions
(Exponential and both Weibull shapes). 
More precisely: 
(i) The capped model overestimates the waste for large platforms (or small MTBFs), in particular for large
values of $\I$ (see Figures~\ref{fig.082.085}(f) and~\ref{fig.04.07}(f)), but this was the 
price to pay for mathematical rigor;  (ii) The uncapped model is accurate for the whole range of the study.
Another striking result is that all strategies taking prediction into account have the same waste as 
their \bestper counterpart, which demonstrates that our formula $\Textr[1] = \sqrt{\frac{2 \mu \C}{1-\recall}}$
is indeed the best possible checkpointing period in regular mode.

Unsurprisingly, \ExactPrediction is better than the heuristics that use a time window instead of exact 
prediction dates, especially with a high 
number of processors. However, interval based heuristics achieve close results when
$\I=300s$, or when $\I=3,000s$ and a small number of processors ($N<2^{16}$).

\begin{figure*}
\centering
\hspace{-1cm}
\subfloat[Exponential]
{
\includegraphics[scale=0.4]{fig/simulations/r085p083I300UNIF-EXP-fixedC-appli0-platform-variation.fig}
}	
\subfloat[Weilbull $k=0.7$]
{
\includegraphics[scale=0.4]{fig/simulations/r085p083I300UNIF-WEIBULL-07-fixedC-appli0-platform-variation.fig}
}
\subfloat[Weilbull $k=0.5$]
{
\includegraphics[scale=0.4]{fig/simulations/r085p083I300UNIF-WEIBULL-05-fixedC-appli0-platform-variation.fig}
}
\\
\hspace{-1cm}
\subfloat[Exponential]
{
\includegraphics[scale=0.4]{fig/simulations/r085p083I3000UNIF-EXP-fixedC-appli0-platform-variation.fig}
}	
\subfloat[Weilbull $k=0.7$]
{
\includegraphics[scale=0.4]{fig/simulations/r085p083I3000UNIF-WEIBULL-07-fixedC-appli0-platform-variation.fig}
}
\subfloat[Weibull $k=0.5$]
{
\includegraphics[scale=0.4]{fig/simulations/r085p083I3000UNIF-WEIBULL-05-fixedC-appli0-platform-variation.fig}
}
\caption{Waste for the different heuristics, with\,$\precision=0.82$,\,$\recall=0.85$,\,$\I=300$s\,(first row)\,or\,$\I=3,000$s\,(second\,row) 
and with a trace of false predictions parametrized by a uniform distribution.}
	\label{fig.082.085.UNIF}
\end{figure*}

\begin{figure*}
\centering
\resizebox{\textwidth}{!}{
\subfloat[Capped periods]
{
\includegraphics[scale=0.16]{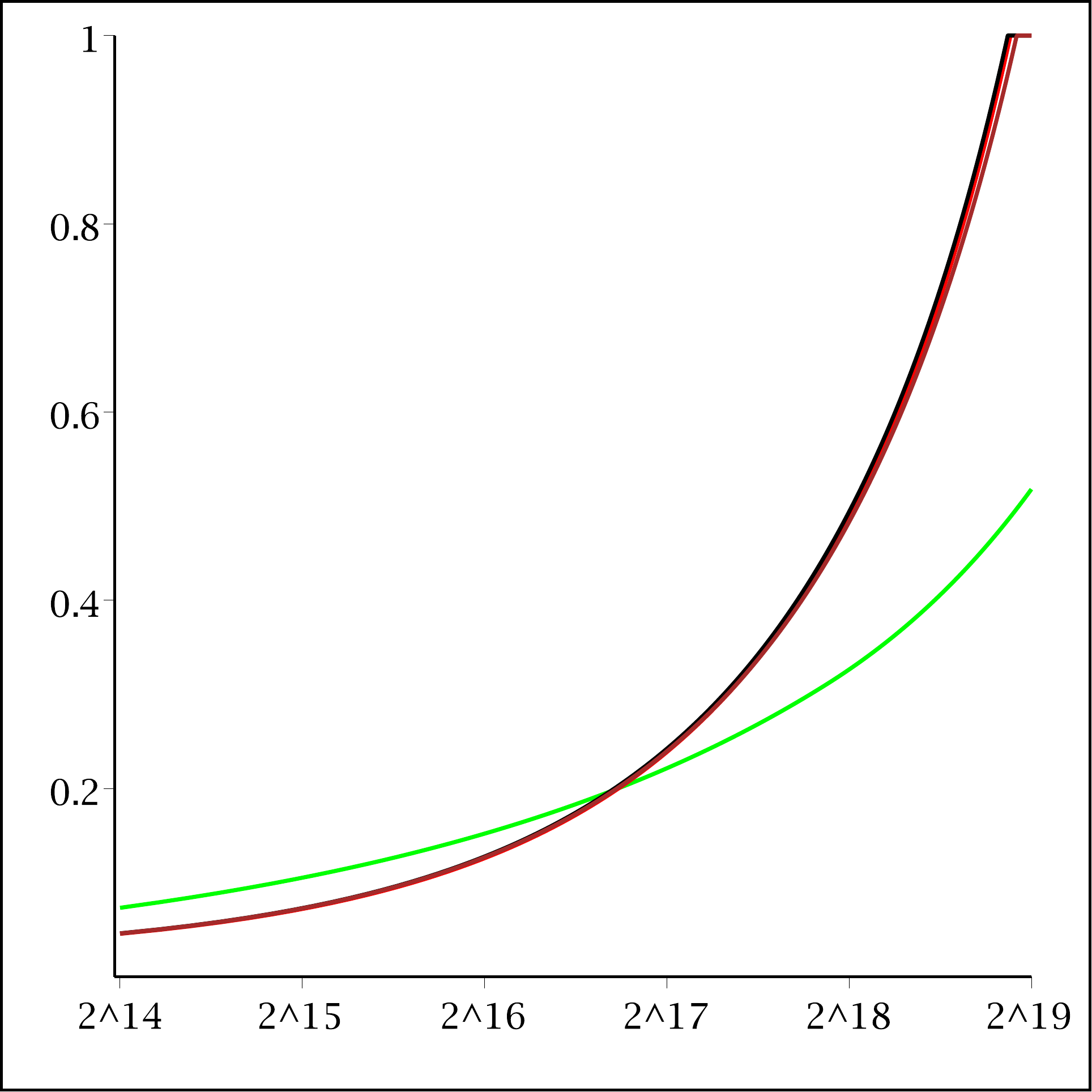}
}
\subfloat[Uncapped periods]
{
\includegraphics[scale=0.16]{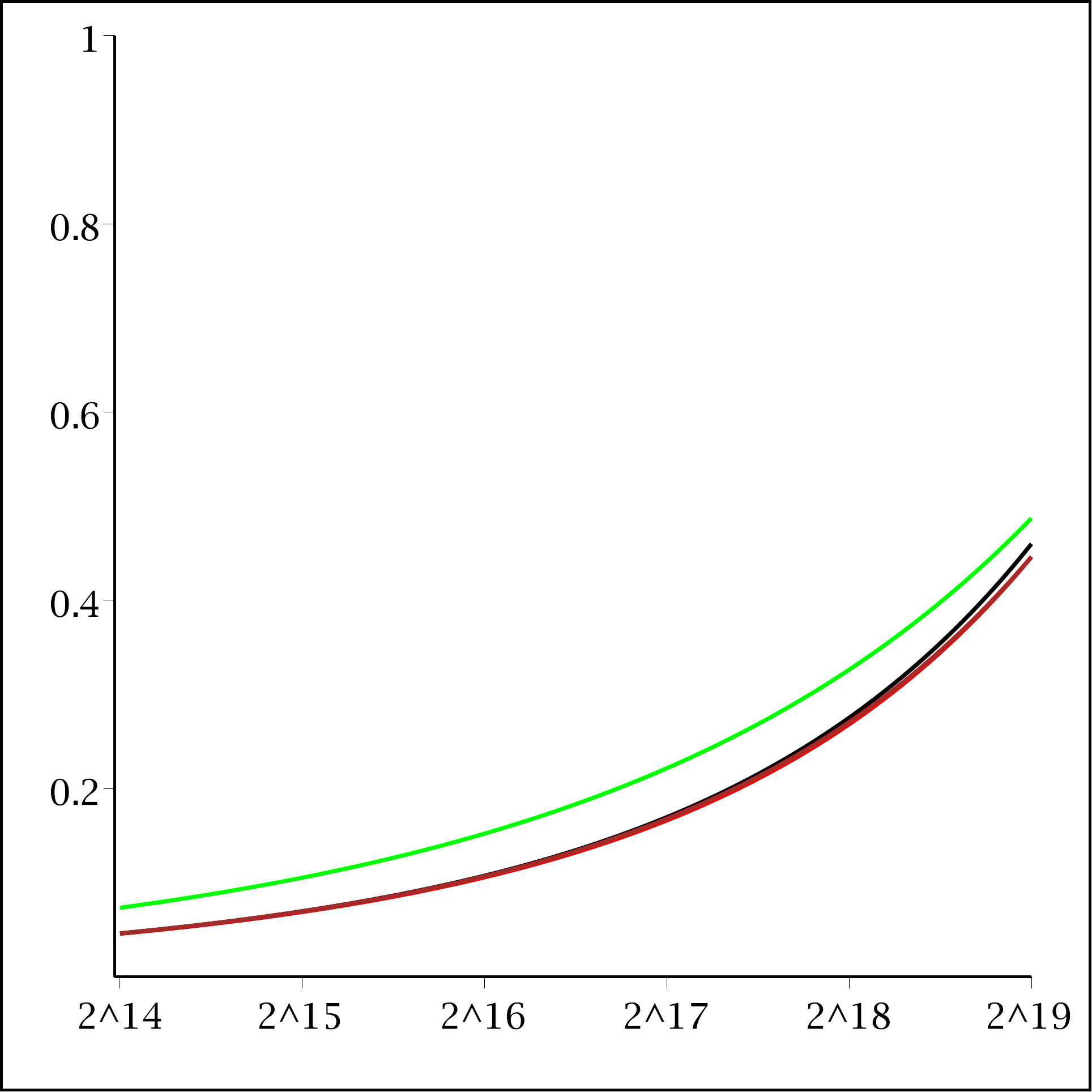}
}
\subfloat[Exponential]
{
\includegraphics[scale=0.36]{fig/simulations/recall07precision04I300-EXP-fixedC-appli0-platform-variation.fig}
}
\subfloat[Weibull $k=0.7$]
{
\includegraphics[scale=0.36]{fig/simulations/recall07precision04I300-WEIBULL-07-fixedC-appli0-platform-variation.fig}
}
\subfloat[Weibull $k=0.5$]
{
\includegraphics[scale=0.36]{fig/simulations/recall07precision04I300-WEIBULL-05-fixedC-appli0-platform-variation.fig}
}
}\\ \resizebox{\textwidth}{!}{
\subfloat[Capped periods]
{
\includegraphics[scale=0.16]{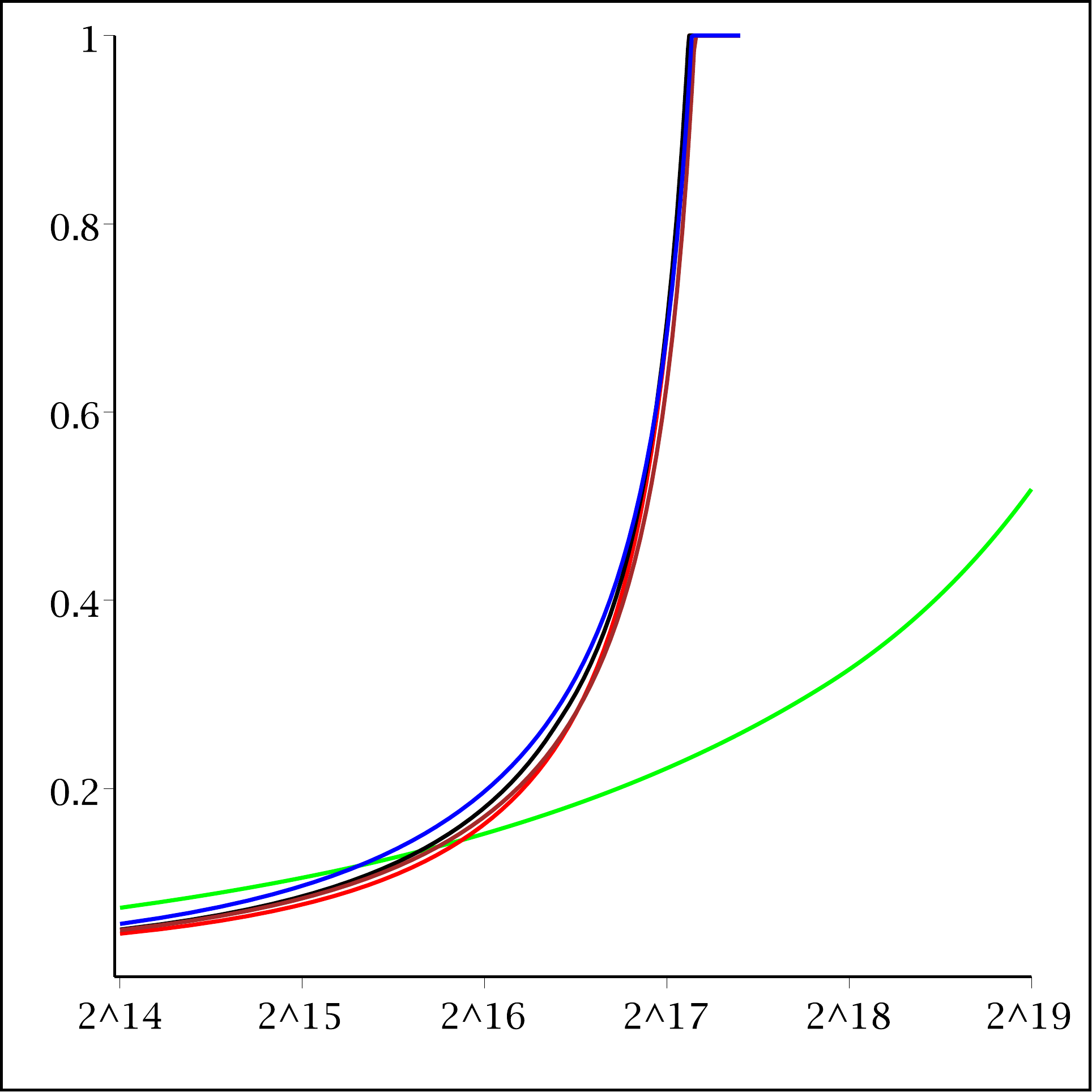}
}
\subfloat[Uncapped periods]
{
\includegraphics[scale=0.16]{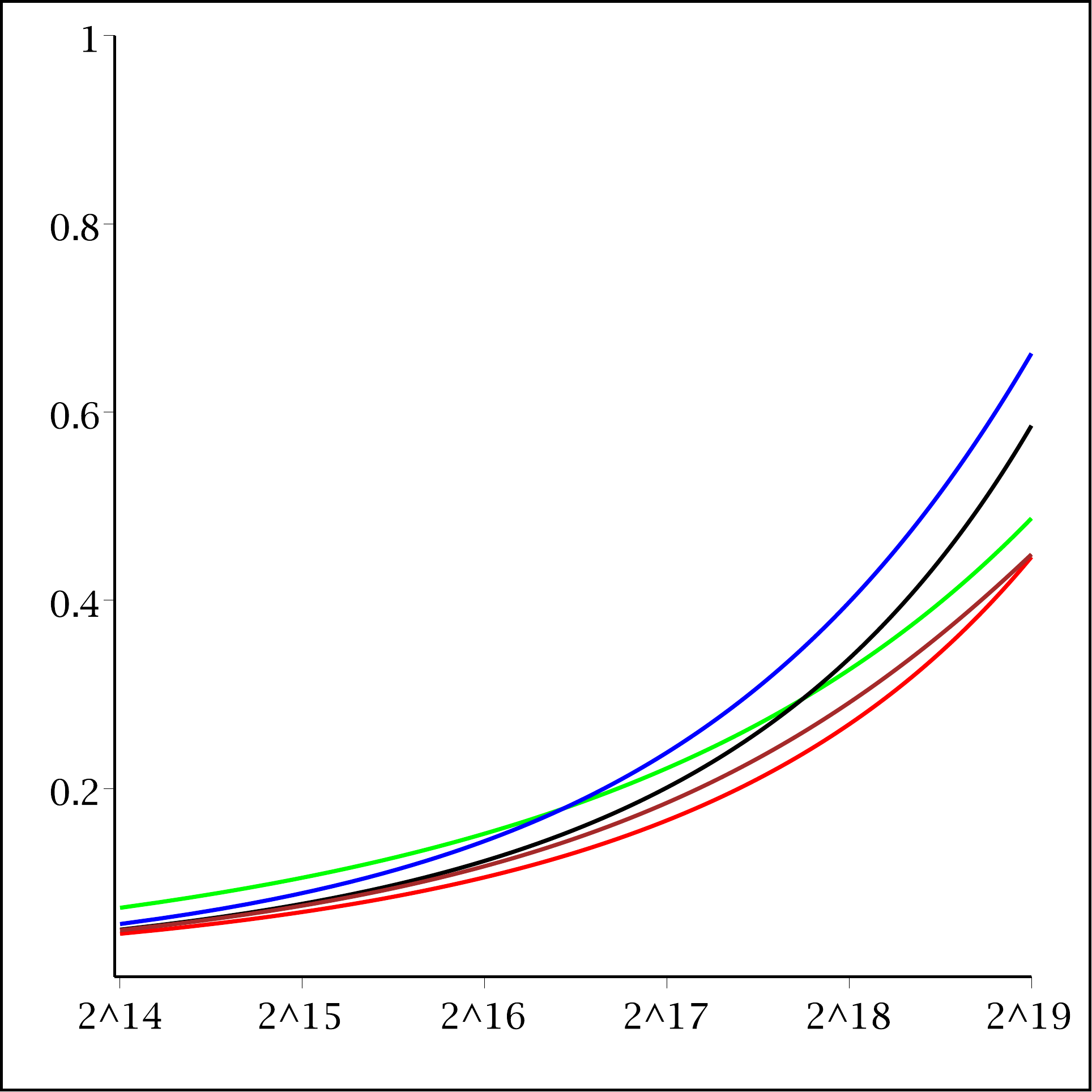}
}
\subfloat[Exponential]
{
\includegraphics[scale=0.36]{fig/simulations/recall07precision04I3000-EXP-fixedC-appli0-platform-variation.fig}
}	
\subfloat[Weibull $k=0.7$]
{
\includegraphics[scale=0.36]{fig/simulations/recall07precision04I3000-WEIBULL-07-fixedC-appli0-platform-variation.fig}
}
\subfloat[Weibull $k=0.5$]
{
\includegraphics[scale=0.36]{fig/simulations/recall07precision04I3000-WEIBULL-05-fixedC-appli0-platform-variation.fig}
}}
\caption{Waste for the different heuristics, with\,$\precision=0.4$,\,$\recall=0.7$,\,$\I=300$s\,(first row)\,or\,$\I=3,000$s\,(second\,row)
and with a trace of false predictions parametrized by a distribution identical to the distribution of the trace of failures.}
	\label{fig.04.07}
\end{figure*}

\begin{figure*}
\centering
\hspace{-1cm}
\subfloat[Exponential]
{
\includegraphics[scale=0.4]{fig/simulations/recall07precision04I300UNIF-EXP-fixedC-appli0-platform-variation.fig}
}
\subfloat[Weibull $k=0.7$]
{
\includegraphics[scale=0.4]{fig/simulations/recall07precision04I300UNIF-WEIBULL-07-fixedC-appli0-platform-variation.fig}
}
\subfloat[Weibull $k=0.5$]
{
\includegraphics[scale=0.4]{fig/simulations/recall07precision04I300UNIF-WEIBULL-05-fixedC-appli0-platform-variation.fig}
}
\\
\hspace{-1cm}
\subfloat[Exponential]
{
\includegraphics[scale=0.4]{fig/simulations/recall07precision04I3000UNIF-EXP-fixedC-appli0-platform-variation.fig}
}	
\subfloat[Weibull $k=0.7$]
{
\includegraphics[scale=0.4]{fig/simulations/recall07precision04I3000UNIF-WEIBULL-07-fixedC-appli0-platform-variation.fig}
}
\subfloat[Weibull $k=0.5$]
{
\includegraphics[scale=0.4]{fig/simulations/recall07precision04I3000UNIF-WEIBULL-05-fixedC-appli0-platform-variation.fig}
}
\caption{Waste for the different heuristics, with\,$\precision=0.4$,\,$\recall=0.7$,\,$\I=300$s\,(first row)\,or\,$\I=3,000$s\,(second\,row) 
and with a trace of false predictions parametrized by a uniform distribution.}
	\label{fig.04.07.UNIF}
\end{figure*}

In order to compare the heuristics without prediction to those with prediction, we report job execution times
in Table~\ref{makespan.300.tab}. For the strategies with prediction, we compute the gain (expressed in percentage)
over \young, the reference strategy without prediction. 
For $\I=300s$, the three strategies are identical. But for $\I=3,000s$, \Withckpt has often better results.
First, with $\precision=0.85$ and $\recall=0.82$ and  
$\I=3,000s$, we save $25\% $ of the total time with $N=2^{19}$, and $14\%$ with $N=2^{16}$
using strategy \Withckpt.
With $\I=300s$, we save up to $44\%$ with $N=2^{19}$, and $18\%$ with $N=2^{16}$
using any strategy (though \Nockpt is slightly better than \Instant).
Then, with $\precision=0.4$ and $\recall = 0.7$, we still save $32\%$ of the execution 
time when $\I=300s$ and $N=2^{19}$, and $13\%$ with $N=2^{16}$.
The gain gets  smaller with $\I=3,000s$, but remains non negligible since we can save up to $9.7\%$
with $N=2^{19}$, and $7.6\%$ with $N=2^{16}$. 
Unexpectedly in this last case, the strategy that is the most efficient is \Instant
and not \Withckpt.

We observe that the size of the prediction-window \I plays an important role too: we have better 
results for $\I=300$ and $(\precision,\recall)=(0.4,0.7)$, than for $\I=3000$ and 
$(\precision,\recall)=(0.82,0.85)$.

In Table~\ref{makespan.300.tab}, we report the job execution times for Weibull distributions with $k=0.5$.

For $\I=300s$, the three strategies are identical. But for $\I=3,000s$, \Withckpt has often better results.
First, with $\precision=0.85$ and $\recall=0.82$ and  
$\I=3,000s$, we save $61\% $ of the total time with $N=2^{19}$, and $30\%$ with $N=2^{16}$
using strategy \Withckpt.

With $\I=300s$, we save up to $74\%$ with $N=2^{19}$, and $38\%$ with $N=2^{16}$
using any strategy (though \Nockpt is slightly better than \Instant).

Then, with $\precision=0.4$ and $\recall = 0.7$, we still save $66\%$ of the execution 
time when $\I=300s$ and $N=2^{19}$, and $33\%$ with $N=2^{16}$.
The gain gets  smaller with $\I=3,000s$, but we can save up to $52\%$
with $N=2^{19}$, and $22\%$ with $N=2^{16}$. 
  
Using a Weibull failure distribution with shape parameter 0.5, we observe that the gain due to prediction is twice larger than the gain computed with a Weibull failure distribution with shape parameter 0.7.
We can conclude the same remark from Figures~\ref{fig.082.085}(e),~\ref{fig.082.085}(j),~\ref{fig.04.07}(e) and~\ref{fig.04.07}(j).

We also performed simulations with a trace of false predictions parametrized by a uniform distribution and we observe that the result (Figures ~\ref{fig.082.085.UNIF} and ~\ref{fig.04.07.UNIF}) are similar to the result (Figures ~\ref{fig.082.085} and ~\ref{fig.04.07}) with simulations with a trace of false predictions parametrized by a distribution identical to the distribution of the trace of failures.

\begin{table}
\centering
\begin{tabular}{c||c|c||c|c||}
 & \multicolumn{2}{|c||}{Execution time (in days)} & \multicolumn{2}{|c||}{Execution time (in days)}\\
$\I=300$  & \multicolumn{2}{|c||}{($\precision = 0.82$,$\recall=0.85$)} & \multicolumn{2}{|c||}{($\precision = 0.4$,$\recall=0.7$)}\\
 & $2^{16}$ procs & $2^{19}$ procs & $2^{16}$ procs & $2^{19}$ procs \\
\hline
\young & 81.3 & 30.1 & 81.2 & 30.1\\
\hline
\ExactPrediction & 65.9 (19\%) & 15.9 (47\%)& 69.7 (14\%) & 19.3 (36\%)\\
\Nockpt & 66.5 (18\%)& 16.9 (44\%)& 70.3 (13\%)& 20.5 (32\%)\\
\Instant & 66.5 (18\%)& 17.0 (44\%)& 70.3 (13\%)& 20.7 (31\%)\\
\end{tabular}
\\
\vspace*{0.2cm}
\begin{tabular}{c||c|c||c|c||}
 & \multicolumn{2}{|c||}{Execution time (in days)} & \multicolumn{2}{|c||}{Execution time (in days)}\\
$\I=3,000$  & \multicolumn{2}{|c||}{($\precision = 0.82$,$\recall=0.85$)} & \multicolumn{2}{|c||}{($\precision = 0.4$,$\recall=0.7$)}\\
 & $2^{16}$ procs & $2^{19}$ procs & $2^{16}$ procs & $2^{19}$ procs \\
\hline
\young & 81.2 & 30.1 & 81.2 & 30.1 \\
\hline
\ExactPrediction & 66.0 (19\%) & 15.9 (47\%)& 69.8 (14\%) & 19.3 (36\%)\\
\Nockpt & 71.1 (12\%) & 24.6 (18\%)& 75.2 (7.3\%)& 28.9 (4.0\%)\\
\Withckpt  & 70.0 (14\%)& 22.6 (25\%)& 75.4 (7.1\%)& 27.2 (9.7\%) \\
\Instant & 71.2 (12\%)& 24.2 (20\%)& 75.0 (7.6\%)& 28.3 (6.0\%)\\
\end{tabular}

\caption{Comparing job execution times for a Weibull distribution ($k=0.7$), and reporting the gain when comparing to \young.}
\label{makespan.300.tab}
\end{table}

\begin{table}
\centering
\begin{tabular}{c||c|c||c|c||}
 & \multicolumn{2}{|c||}{Execution time (in days)} & \multicolumn{2}{|c||}{Execution time (in days)}\\
$\I=300$  & \multicolumn{2}{|c||}{($\precision = 0.82$,$\recall=0.85$)} & \multicolumn{2}{|c||}{($\precision = 0.4$,$\recall=0.7$)}\\
 & $2^{16}$ procs & $2^{19}$ procs & $2^{16}$ procs & $2^{19}$ procs \\
\hline
\young & 125.4 & 171.8 & 125.5 & 171.7\\
\hline
\ExactPrediction & 75.8 (40\%) & 39.4 (77\%)& 82.9 (34\%) & 51.8(70\%)\\
\Nockpt & 77.3 (38\%)& 44.8 (74\%)& 84.6 (33\%)& 58.2 (66\%)\\
\Instant & 77.4 (38\%)& 45.1 (74\%)& 84.7 (33\%)& 59.1 (66\%)\\
\end{tabular}
\\
\vspace*{0.2cm}
\begin{tabular}{c||c|c||c|c||}
 & \multicolumn{2}{|c||}{Execution time (in days)} & \multicolumn{2}{|c||}{Execution time (in days)}\\
$\I=3,000$  & \multicolumn{2}{|c||}{($\precision = 0.82$,$\recall=0.85$)} & \multicolumn{2}{|c||}{($\precision = 0.4$,$\recall=0.7$)}\\
 & $2^{16}$ procs & $2^{19}$ procs & $2^{16}$ procs & $2^{19}$ procs \\
\hline
\young & 125.4 & 171.9 & 125.4 & 172.0 \\
\hline
\ExactPrediction & 76.1 (39\%) & 39.4 (77\%)& 83.0 (34\%) & 51.7 (70\%)\\
\Nockpt & 90.0 (28\%) &  71.8 (58\%)& 98.3 (22\%)& 84.5 (51\%)\\
\Withckpt  & 87.8 (30\%)& 66.6 (61\%)& 98.0 (22\%)& 82.2 (52\%) \\
\Instant & 89.8 (28\%)& 70.9 (59\%)& 98.2 (22\%)& 83.2 (52\%)\\
\end{tabular}

\caption{Comparing job execution times for a Weibull distribution ($k=0.5$), 
and reporting the gain when comparing to \young.}
\label{makespan.300.tab}
\end{table}

\subsection{Recall vs. precision}
\label{sec.impact}

In this section, we assess the impact of the two key parameters of the predictor,  its recall \recall and its precision $\precision$.
To this purpose, we conduct simulations where one parameter is fixed, and we let the other vary. We choose two platforms,
a smaller one with  $N=2^{16}$ processors (or a MTBF $\mu=1,000mn$) and the other with
$N=2^{19}$ processors (or a MTBF $\mu=125mn$).  In both cases, we use a prediction-window of size $\I=300s$,
and a Weibull failure distribution with shape parameter $k=0.7$ (we have similar results (Figures ~\ref{fig.recall.19.05} and ~\ref{fig.precision.19.05}) for $k=0.5$).

\begin{figure*}
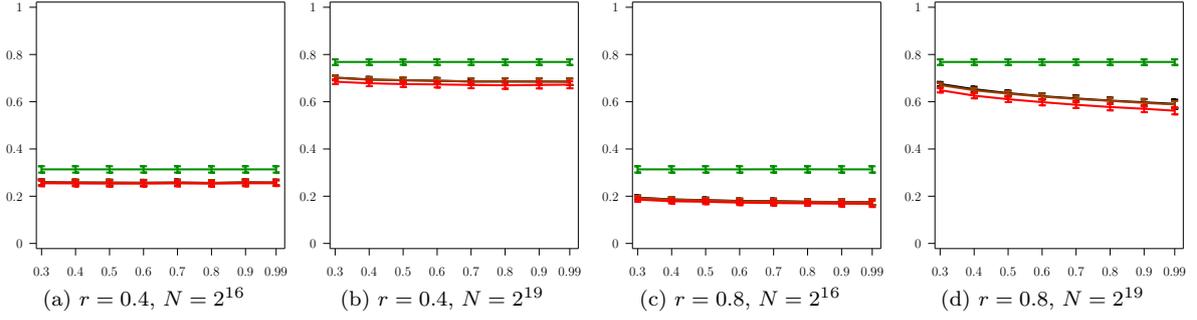

\centering
\resizebox{\textwidth}{!}{
\subfloat[$\recall=0.4$, $N=2^{16}$]
{
\includegraphics[scale=0.41]{fig/simulations/recall04I300platform16-WEIBULL-07-fixedC-appli0-platform-variation.fig}
}
\subfloat[$\recall=0.4$, $N=2^{19}$]
{
\includegraphics[scale=0.41]{fig/simulations/recall04I300platform19-WEIBULL-07-fixedC-appli0-platform-variation.fig}
}
\subfloat[$\recall=0.8$, $N=2^{16}$]
{
\includegraphics[scale=0.41]{fig/simulations/recall08I300platform16-WEIBULL-07-fixedC-appli0-platform-variation.fig}
}
\subfloat[$\recall=0.8$, $N=2^{19}$]
{
\includegraphics[scale=0.41]{fig/simulations/recall08I300platform19-WEIBULL-07-fixedC-appli0-platform-variation.fig}
}
}
\caption{Impact of the precision for a fixed recall ($\recall=0.4$ and  $\recall=0.8$) and for a Weibull distribution (k=0.7).}
	\label{fig.recall.19}
\end{figure*}

In Figure~\ref{fig.recall.19}, we fix the value of \recall (either $\recall=0.4$ or $\recall=0.8$) and we let 
$\precision$ vary from $0.3$ to $0.99$.  In the four plots, we observe that the precision has a minor impact on
the waste.
In Figure~\ref{fig.precision.19}, we conduct the opposite experiment and  fix the value of \precision (either $ \precision=0.4$
or  $\precision=0.8$), letting $\recall$ vary from $0.3$ to $0.99$.  Here we observe that increasing the recall can 
significantly improve the performance.

Altogether we conclude that it is more important (for the design of future predictors) 
 to focus on improving the recall \recall rather than the precision \precision, and our results can help quantify this statement.
 We provide an intuitive explanation as follows: unpredicted failures prove very harmful and heavily increase the waste, 
 while unduly checkpointing due to false predictions turns out to induce a smaller overhead.
 
\begin{figure*}
\centering
\resizebox{\textwidth}{!}{
\subfloat[$\recall=0.4$, $N=2^{16}$]
{
\includegraphics[scale=0.41]{fig/simulations/recall04I300platform16-WEIBULL-05-fixedC-appli0-platform-variation.fig}
}
\subfloat[$\recall=0.4$, $N=2^{19}$]
{
\includegraphics[scale=0.41]{fig/simulations/recall04I300platform19-WEIBULL-05-fixedC-appli0-platform-variation.fig}
}
\subfloat[$\recall=0.8$, $N=2^{16}$]
{
\includegraphics[scale=0.41]{fig/simulations/recall08I300platform16-WEIBULL-05-fixedC-appli0-platform-variation.fig}
}
\subfloat[$\recall=0.8$, $N=2^{19}$]
{
\includegraphics[scale=0.41]{fig/simulations/recall08I300platform19-WEIBULL-05-fixedC-appli0-platform-variation.fig}
}
}
\caption{Impact of the precision for a fixed recall ($\recall=0.4$ and  $\recall=0.8$) and for a Weibull distribution (k=0.5).}
	\label{fig.recall.19.05}
\end{figure*}

\begin{figure*}
\centering
\resizebox{\textwidth}{!}{
\subfloat[$\precision=0.4$, $N=2^{16}$]
{
\includegraphics[scale=0.41]{fig/simulations/precision04I300platform16-WEIBULL-07-fixedC-appli0-platform-variation.fig}
}
\subfloat[$\precision=0.4$, $N=2^{19}$]
{
\includegraphics[scale=0.41]{fig/simulations/precision04I300platform19-WEIBULL-07-fixedC-appli0-platform-variation.fig}
}
\subfloat[$\precision=0.8$, $N = 2^{16}$]
{
\includegraphics[scale=0.41]{fig/simulations/precision08I300platform16-WEIBULL-07-fixedC-appli0-platform-variation.fig}
}
\subfloat[$\precision=0.8$, $N=2^{19}$]
{
\includegraphics[scale=0.41]{fig/simulations/precision08I300platform19-WEIBULL-07-fixedC-appli0-platform-variation.fig}
}
}
\caption{Impact of the recall for a fixed precision ($\precision=0.4$ and  $\precision=0.8$) and for a Weibull distribution (k=0.7).}
	\label{fig.precision.19}
\end{figure*}

\begin{figure*}
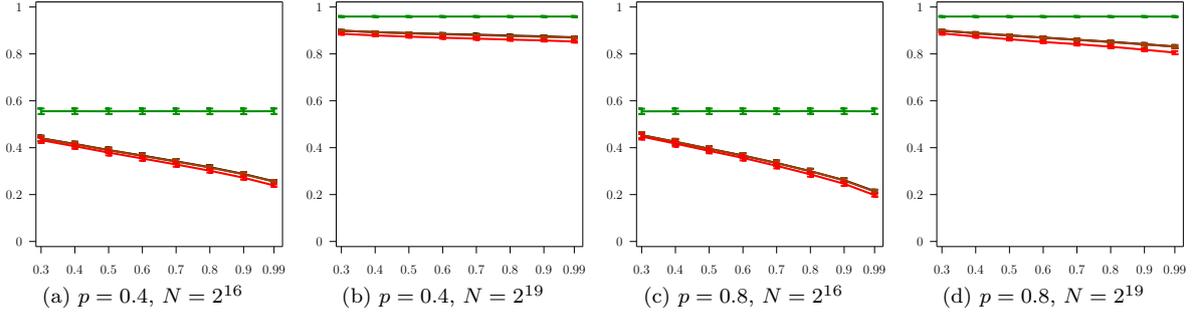

\centering
\resizebox{\textwidth}{!}{
\subfloat[$\precision=0.4$, $N=2^{16}$]
{
\includegraphics[scale=0.41]{fig/simulations/precision04I300platform16-WEIBULL-05-fixedC-appli0-platform-variation.fig}
}
\subfloat[$\precision=0.4$, $N=2^{19}$]
{
\includegraphics[scale=0.41]{fig/simulations/precision04I300platform19-WEIBULL-05-fixedC-appli0-platform-variation.fig}
}
\subfloat[$\precision=0.8$, $N = 2^{16}$]
{
\includegraphics[scale=0.41]{fig/simulations/precision08I300platform16-WEIBULL-05-fixedC-appli0-platform-variation.fig}
}
\subfloat[$\precision=0.8$, $N=2^{19}$]
{
\includegraphics[scale=0.41]{fig/simulations/precision08I300platform19-WEIBULL-05-fixedC-appli0-platform-variation.fig}
}
}
\caption{Impact of the recall for a fixed precision ($\precision=0.4$ and  $\precision=0.8$) and for a Weibull distribution (k=0.5). }
	\label{fig.precision.19.05}
\end{figure*}

\section{Related work}
\label{sec.related}

Considerable research has been conducted on fault prediction using different models (system log
analysis~\cite{5958823}, event-driven approach~\cite{GainaruIPDPS12,5958823,5542627}, 
support vector machines~\cite{LiangZXS07,Fulp:2008:PCS:1855886.1855891}), nearest neighbors~\cite{LiangZXS07}, \dots).
In this section we give a brief overview of the results obtained by predictors. We focus on their 
results rather than on their methods of prediction.

The authors of~\cite{5542627} introduce the \emph{lead time}, that is the time between the prediction and the 
actual fault. This time should be sufficient to take proactive actions. They are also able to give 
the location of the fault. While this has a negative impact on the precision (see the low value of 
\precision in Table~\ref{rel.work.tab}), they state that it has a positive impact on the checkpointing
time (from 1500 seconds to 120 seconds).
The authors of~\cite{5958823} also consider a lead time, and  introduce a \emph{prediction window} when 
the predicted fault should happen. The authors of~\cite{LiangZXS07} study the impact of different prediction techniques with different 
prediction window sizes. They also consider a lead time, but do not state its value.
These two latter studies motivate the work of Section~\ref{sec.intervals}, 
even though~\cite{5958823} does not provide the size of their prediction window.

\begin{table}
\centering
\resizebox{\linewidth}{!}{%
\begin{tabular}{|c|c|c|c|c|}
Paper & Lead Time & Precision & Recall & Prediction Window \\
\hline
\cite{5542627} & 300 s & 40 \% & 70\% & - \\
\cite{5542627} & 600 s & 35 \% & 60\% & - \\
\cite{5958823} & 2h & 64.8 \% & 65.2\% & yes (size unknown) \\
\cite{5958823} & 0 min & 82.3 \% & 85.4 \% & yes (size unknown) \\
\cite{GainaruIPDPS12} & 32 s & 93 \% & 43 \% & - \\
\cite{Fulp:2008:PCS:1855886.1855891} & NA & 70 \% & 75 \% & - \\
\cite{LiangZXS07} & NA & 20 \% & 30 \%& 1h \\
\cite{LiangZXS07} & NA & 30 \%& 75 \% & 4h \\
\cite{LiangZXS07} & NA & 40 \%& 90 \% & 6h \\
\cite{LiangZXS07} & NA & 50 \% & 30 \% & 6h \\
\cite{LiangZXS07} & NA & 60 \% & 85\% & 12h \\
\end{tabular}}
\caption{Comparative study of different parameters returned by some predictors.}
\label{rel.work.tab}
\end{table}

Unfortunately, much of the work done on prediction does not provide information that could be 
really useful for the design of efficient algorithms. These informations are those stated above, namely the lead time 
and the size of the prediction window, but other information that could be useful would be: (i) the 
distribution of the faults in the prediction window; (ii) the precision as a function of the recall (see our 
analysis); and (iii) the precision and recall as functions of the prediction window (what happens with 
a larger prediction window). 

While many studies on fault prediction focus on the conception of the predictor, most of them 
consider that the proactive action should simply be a checkpoint or a migration right in time before the fault. 
However, in their paper~\cite{Fu:2007:EEC:1362622.1362678}, Li et al. consider the 
mathematical problem to determine when and how to migrate. In order to be able to use migration, 
they stated that at every time, 2\% of the resources are available. This allowed them to conceive a 
Knapsack-based heuristic. Thanks to their algorithm, they were able to save 30\% of the execution 
time compared to an heuristic that does not take the reliability into account, with a precision and 
recall of 70\%, and with a maximum load of 0.7.

Finally, to the best of our knowledge, this work is the first to focus on the mathematical aspect of 
fault prediction, and to provide a model and a detailed analysis of the waste due to all three types of
events (true and false predictions and unpredicted failures).

\section{Conclusion}
\label{sec.conclusion}

In this work, we have studied the impact of prediction, either with exact dates or window-based, 
on checkpointing strategies. 
We have designed several algorithms that decide when to trust these predictions, and when it is 
worth taking preventive checkpoints. We have introduced an analytical model to capture the waste incurred by each
strategy, and provided the optimal solution to the corresponding optimization problems.
We have been able to derive some striking conclusions:\\
$\bullet$ The model is quite accurate and its validity goes beyond the conservative assumption that requires capping
checkpointing periods to diminish the probability of having several faults within the same period;\\
$\bullet$ A unified formula for the optimal checkpointing period is $\sqrt{ \dfrac{2 \mu\C }{1-\recall \trust}}$, which unifies both cases
with and without prediction, and nicely extends the work of Young and Daly to account for prediction;\\
$\bullet$ The simulations fully validate the model,  and show that: (i) A significant gain is induced by using predictions,
even for mid-range values of recall and precision; and (ii) The best period (found by brute-force search) is 
always very close to the one predicted by the model and given by the
previous unified formula; this holds true both for Exponential and Weibull failure distributions;\\
$\bullet$ The recall has more impact on the waste than the precision: \emph{better safe than sorry}, or better prepare for a false event than miss an actual failure!

Altogether, the analytical model and the comprehensive results provided in this work enable to
fully assess the impact of fault prediction on optimal checkpointing strategies. 
Future
work will be devoted to refine the assessment of the usefulness of prediction with trace-based failure and prediction logs
from current large-scale supercomputers.

\bigskip
\noindent{\em Acknowledgments.} 
The authors are with Universit\'e de Lyon, France.
Y.~Robert is with the Institut Universitaire de France.
This work was supported in part by the ANR {\em RESCUE} project.

\bigskip
\bibliographystyle{plain}
\bibliography{biblio}

\end{document}